\documentclass[sigconf]{acmart}

\usepackage{booktabs} % For formal tables
\usepackage{xcite}
\usepackage{xr}
\usepackage{graphicx}
\usepackage{caption}
\usepackage{subcaption}
\usepackage{multirow}
\usepackage{bigstrut}
\usepackage{enumitem}
\usepackage[skip=0pt]{caption}
\usepackage[ruled,linesnumbered]{algorithm2e}
\usepackage{verbatim}

\externalcitedocument{introduction}
\externalcitedocument{methods}
\externalcitedocument{evaluation}
\renewcommand\footnotetextcopyrightpermission[1]{} % removes footnote with conference information in first column
\makeatletter
\def\subsubsection{\@startsection{subsubsection}{3}%
  \z@{.3\linespacing\@plus.7\linespacing}{.1\linespacing}%
  {\normalfont\itshape}}
\makeatother

\begin{document}
\title{Point-of-Interest Recommendation: Exploiting Self-Attentive Autoencoders with Neighbor-Aware Influence}

\author{Chen Ma}
\affiliation{\institution{McGill University}}
\email{chen.ma2@mail.mcgill.ca}

\author{Yingxue Zhang}
\affiliation{\institution{McGill University}}
\email{yingxue.zhang2@mail.mcgill.ca}

\author{Qinglong Wang}
\affiliation{\institution{McGill University}}
\email{qinglong.wang@mail.mcgill.ca}

\author{Xue Liu}
\affiliation{\institution{McGill University}}
\email{xueliu@cs.mcgill.ca}

\begin{abstract}
The rapid growth of Location-based Social Networks (LBSNs) provides a great opportunity to satisfy the strong demand for personalized Point-of-Interest (POI) recommendation services. However, with the tremendous increase of users and POIs, POI recommender systems still face several challenging problems: (1) the hardness of modeling non-linear user-POI interactions from implicit feedback; (2) the difficulty of incorporating context information such as POIs' geographical coordinates. To cope with these challenges, we propose a novel autoencoder-based model to learn the non-linear user-POI relations, namely \textit{SAE-NAD}, which consists of a self-attentive encoder (SAE) and a neighbor-aware decoder (NAD). In particular, unlike previous works equally treat users' checked-in POIs, our self-attentive encoder adaptively differentiates the user preference degrees in multiple aspects, by adopting a multi-dimensional attention mechanism. To incorporate the geographical context information, we propose a neighbor-aware decoder to make users' reachability higher on the similar and nearby neighbors of checked-in POIs, which is achieved by the inner product of POI embeddings together with the radial basis function (RBF) kernel. To evaluate the proposed model, we conduct extensive experiments on three real-world datasets with many state-of-the-art baseline methods and evaluation metrics. The experimental results demonstrate the effectiveness of our model.
\end{abstract}

% We no longer use \terms command
%\terms{Theory}

%\keywords{Point-of-Interest; Recommendation; Embedding Representation; Matrix Factorization}

\maketitle

\section{Introduction}
With the rapid growth of mobile devices and location-acquisition technologies, it has become more convenient for people to access their real-time location information. This development enables the advent of Location-based Social Networks (LBSNs), such as Yelp and Foursquare. These LBSNs allow users to connect with each other, post physical positions and share experiences associated with a location, namely, Point-of-Interest (POI). The large amount of user-POI interaction data facilitates a promising service---personalized POI recommendation. POI recommender systems serve a potentially huge service demand and bring significant benefits to at least two parties: (1) help residents or tourists to explore interesting unvisited places; (2) create opportunities for POIs to attract more visitors.

In the literature, effective methods have been proposed for personalized POI recommendation. These methods mainly rely on collaborative filtering (CF), which can be divided into memory-based and model-based methods \cite{DBLP:journals/kbs/BobadillaOHG13}. Memory-based methods infer a user's preferences regarding unvisited POIs based on the weighted average of ratings from similar users or POIs. For example, \cite{DBLP:conf/sigir/YeYLL11} and \cite{DBLP:conf/gis/ZhangC13} applied friend-based collaborative filtering methods to recommend POIs by the similarities between a user and her friends. On the other hand, model-based methods make use of the collection of user-POI records to learn a model for the recommendation. Popularized by the Netflix Prize, some of the most successful realizations of model-based methods are built on matrix factorization (MF). MF discovers the latent features underlying the interactions between users and POIs, which predicts user preferences by the inner product of latent factors. For instance, \cite{DBLP:conf/kdd/LianZXSCR14}, \cite{DBLP:conf/cikm/LiuWSM14} and \cite{DBLP:conf/kdd/LiGHZ16} adopted a weighted regularized MF to infer user preferences on unvisited POIs. 

%However, the huge increase of users and POIs makes the user-POI interaction data complex and complicated. The aforementioned methods, i.e., memory-based and MF-based, which model user preferences by weighted average of ratings and inner product of latent factors, can only capture the linear relations between users and POIs. And in \cite{DBLP:conf/www/HeLZNHC17}, He et al. further illustrates how the inner product function combines latent features linearly and limits the expressiveness of MF.

However, the aforementioned methods may not fully leverage the complicated user-POI interactions in the large-scale data. They usually model user preferences by the weighted average of ratings or the inner product of latent factors, which can only capture the linear relations between users and POIs. It has been shown in \cite{DBLP:conf/www/HeLZNHC17} that how the inner product combines latent features linearly and limits the expressiveness of MF.

Recently, due to the ability to represent non-linear and complex data, autoencoders have been a great success in the domain of recommendation and bring more opportunities to reshape the conventional recommendation architectures \cite{DBLP:conf/kdd/WangWY15,DBLP:conf/cikm/LiKF15,DBLP:conf/wsdm/WuDZE16}. Motivated by this, we propose an autoencoder-based model to cope with the complicated user-POI check-in data. The primary reason we adopt the stacked autoencoder (AE) is that, with the deep neural network structure and the non-linear activation function, the stacked AE effectively captures the non-linear and non-trivial relationships between users and POIs, and enables more complex data representations in the latent space. Besides, AE also has strong relations with multiple MF methods \cite{DBLP:conf/wsdm/WuDZE16}, which can be directly utilized to model the user rating data. 

Nevertheless, applying AE in POI recommendation is a non-trivial task. There are still several challenges. First, we argue that, in the user check-in records, some POIs are more representative than others to reflect users' preferences. Equally treating these representative POIs along with other POIs may lead to inaccurate understanding of users' preferences. Hence, how to further distinguish the user preference degrees on checked-in POIs is significant for learning personalized user preferences. Second, the spatial context information is a unique property in the check-in records, which is critical for improving recommendation performance. Therefore, how to incorporate the auxiliary information into the neural network-based method is a problem. Third, check-in data is a kind of implicit feedback, which means there are only positive samples in the data, and all of the negative samples and missing positive samples are mixed together \cite{DBLP:conf/icdm/PanZCLLSY08}. Moreover, users can only visit a small number of POIs from millions of POIs, which makes the user-POI check-in data extremely sparse. Thus, how to capture users' preferences from the sparse implicit feedback is a challenge.

To address the challenges above, we propose a novel autoencoder-based model, \textit{SAE-NAD}, which consists of two components: a self-attentive encoder (SAE) and a neighbor-aware decoder (NAD). First, unlike existing methods do not deeply explore the implicitness of users' preferences, we propose the self-attentive encoder to adaptively compute an importance vector for each POI in a user's check-in records, which demonstrates the user preference on those checked-in POIs in multiple aspects. As such, users' preferences on checked-in POIs can be further distinguished. The POIs with larger importance values will contribute more to learn the user hidden representation, which can make the user hidden representation more personalized. Second, we propose the neighbor-aware decoder to incorporate the geographical influence \cite{DBLP:conf/sigir/YeYLL11,DBLP:conf/aaai/ChengYKL12}, which widely exists in the human mobility behavior on LBSNs. We adopt the inner product between the embeddings of checked-in POIs and unvisited POIs, together with the radial basis function (RBF) kernel (based on the pair-wise distance of corresponding POIs), to calculate the influence checked-in POIs applied on unvisited POIs. By doing this, the user reachability on the neighbors that are similar and close to checked-in POIs will be higher than the distant POIs. Third, to model the sparse implicit feedback, we assign the same small weights to all unvisited POIs and assign larger weights to visited POIs according to the visit frequency, which makes a distinction between unvisited POIs, less-visited POIs and frequent-visited POIs for each user. We extensively evaluate our model with many state-of-the-art baseline methods and different validation metrics on three real-world datasets. The experimental results demonstrate the improvements of our model over other baseline methods on POI recommendation.

The major contributions of this paper are as follows:
\begin{itemize}[leftmargin=*]
\item Due to the fact that memory-based and MF-based CF can only capture the linear relationships between users and POIs, we adopt an autoencoder-based model to capture the non-linear and complex user-POI interactions in the check-in data.
\item To distinguish the user preference on checked-in POIs, we propose a self-attentive encoder to adaptively compute an importance vector for each checked-in POI, and make the POI contribute to the user hidden representation according to the importance values. To the best of our knowledge, this is the first paper to use attention-based autoencoders in POI recommendation.
\item To incorporate the geographical influence, we propose a neighbor-aware decoder, which adopts the inner product between the embeddings of checked-in POIs and unvisited POIs, along with the RBF kernel based on the distance of POIs, to model the influence checked-in POIs exerted on unvisited POIs. 
\item The proposed model achieves the best performance on three real-world datasets comparing to the state-of-the-art methods, exhibiting the superiority of our model.
\end{itemize}

\section{Related Work}
POI recommendation, also referred to location recommendation or venue recommendation, is an important topic in the domain of recommender systems \cite{DBLP:journals/geoinformatica/0003ZWM15}. In this section, we describe related work in personalized location recommendation, as well as the applications of attention mechanisms in recommendation tasks.

\subsection{Personalized Location Recommendation}
Recently, with the advance of LBSNs, location recommendation has been widely studied. User's historical data (check-ins, comments, etc.) was used to make the recommendation personalized. Most of the proposed methods, using historical records, are based on collaborative filtering (CF). Some researchers employ memory-based CF \cite{DBLP:journals/kbs/BobadillaOHG13} to learn user preferences \cite{DBLP:conf/gis/YeYL10,DBLP:conf/sigir/YeYLL11}. For example , Ye et al. \cite{DBLP:conf/gis/YeYL10} proposed a friend-based CF integrating the preferences of user's social friends, which was based on user-based CF. On the other hand, recent work utilized model-based CF \cite{DBLP:journals/kbs/BobadillaOHG13} to recommend POIs \cite{DBLP:conf/icdm/LiHZG15,DBLP:conf/aaai/ChengYKL12,DBLP:conf/cikm/YinZSWS15}, such as matrix factorization \cite{DBLP:journals/computer/KorenBV09}. Furthermore, in \cite{DBLP:conf/kdd/LianZXSCR14}, \cite{DBLP:conf/cikm/LiuWSM14} and \cite{DBLP:conf/kdd/LiGHZ16}, researchers found check-ins can be treated as implicit feedback, and applied the weighted regularized matrix factorization \cite{DBLP:conf/icdm/HuKV08} to model the implicit feedback data. While other researchers considered the recommendation problem as a task of pair-wise ranking. In \cite{DBLP:conf/ijcai/ChengYLK13} and \cite{DBLP:conf/aaai/ZhaoZYLK16}, researchers adopted Bayesian personalized ranking loss \cite{DBLP:conf/uai/RendleFGS09} to learn the pair-wise preferences on POIs.

To make more accurate recommendations, researchers incorporated POI geographical influence into their proposed models \cite{DBLP:conf/kdd/ChoML11,DBLP:conf/sigir/YeYLL11,DBLP:conf/aaai/ChengYKL12,DBLP:conf/kdd/LiuFYX13,DBLP:conf/cikm/YuanCS14,DBLP:conf/cikm/LiuWSM14}. There are several ways to model the geographical influence. In particular, some researchers employed Gaussian distribution to characterize user's check-in activities. For example, Cho et al. \cite{DBLP:conf/kdd/ChoML11} applied a two-state Gaussian mixture to model the check-ins that close to users' home or work places. Cheng et al. \cite{DBLP:conf/aaai/ChengYKL12} proposed a multi-center discovering algorithm to detect user's check-in centers. Then Gaussian distribution was built on each center, calculating user check-in probabilities on unvisited locations together. On the other hand, some researchers proposed the kernel density estimation (KDE) to estimate user's check-in activities. Ye et al. \cite{DBLP:conf/sigir/YeYLL11} discovered that user's check-in behaviors were in a power law distribution pattern. The power law pattern revealed two locations' co-occurrence probability distribution over their distance, and this discovery was also employed in \cite{DBLP:conf/kdd/LiuFYX13}. Besides, Liu et al. \cite{DBLP:conf/cikm/LiuWSM14} exploited geographical characteristics from a location perspective, which were modeled by two levels of neighborhoods, i.e., instance-level and region-level.

%Furthermore, social relations are also a significant part in POI recommendation \cite{DBLP:conf/wsdm/MaZLLK11,DBLP:conf/icwsm/GaoTL12,DBLP:conf/sigir/YeYLL11,DBLP:conf/kdd/LiGHZ16,DBLP:conf/ijcai/TangHGL13,DBLP:conf/icdm/LiHZG15}. For example, Ye et al. \cite{DBLP:conf/sigir/YeYLL11} designed a similarity function based on users' social connections and check-in activities. Furthermore, this similarity was incorporated into the friend-based CF. Ma et al. \cite{DBLP:conf/wsdm/MaZLLK11} assumed users would share similar interests with their friends, and put a social regularization term to constrain the objective function. Li et al. \cite{DBLP:conf/kdd/LiGHZ16} generated a user's potential check-in list by combining check-ins of her social friends, location friends and neighboring friends. Tang et al. \cite{DBLP:conf/ijcai/TangHGL13} investigated how to exploit local and global social context for the recommendation. In the local level, they modeled the relations between user and her friends by fitting the similarities on check-in activities. In the global level, they used user reputation scores to weight the importance of their ratings.

Recently, deep neural networks are also applied in POI recommendation. In \cite{DBLP:conf/kdd/YangBZY017}, Yang et al. proposed to exploit context graphs and apply user/POI smoothing to address data sparsity and various context. In \cite{DBLP:conf/cikm/ManotumruksaMO17}, Manotumruksa et al. proposed a deep recurrent collaborative filtering framework with a pairwise ranking function.

% \subsection{Autoencoders for Recommendation}
% With the tremendous development of deep learning, AE has been an important building block and widely used in a number of domains, such as computer vision and natural language processing. Recently, AE has been a popular choice for building recommender systems. On the one hand, AE is used for capturing auxiliary information. In \cite{DBLP:conf/kdd/WangWY15}, Wang et al. proposed a hierarchical Bayesian model called collaborative deep learning (CDL), which couples the stacked denoising autoencoder (SDAE) for the content information, along with collaborative filtering for the rating matrix. Li et al. in \cite{DBLP:conf/cikm/LiKF15} proposed to combine probabilistic matrix factorization with marginalized denoising autoencoders. On the other hand, there are several studies rely solely on AE. Sedhain et al. \cite{DBLP:conf/www/SedhainMSX15} proposed AutoRec for rating prediction, which learns latent representations from users' historical ratings, then reconstructs users' predicted ratings from the latent representations. In \cite{DBLP:conf/wsdm/WuDZE16}, Wu et al. proposed the collaborative denoising autoencoder (CDAE) for top-n recommendation from implicit feedback, which additionally plugs a user node to the input of denoising autoencoders for reconstructing the user's ratings.

\subsection{Attention Mechanism in Recommendation}
The idea of attention mechanism in neural networks is loosely based on the visual attention found in humans, which has demonstrated the effectiveness in various machine learning tasks such as document classification \cite{DBLP:conf/naacl/YangYDHSH16} and machine translation \cite{DBLP:conf/emnlp/LuongPM15,DBLP:conf/nips/VaswaniSPUJGKP17}. 

Recently, researchers also adopt attention mechanism on recommendation tasks. In \cite{DBLP:conf/cikm/PeiYSZBT17}, Pei et al. adopted an attention model to measure the relevance between users and items, which can capture the joint effects on user-item interactions. Wang et al. \cite{DBLP:conf/kdd/WangYRTZYW17} proposed a hybrid attention model to adaptively capture the change of editors' selection criteria. In \cite{DBLP:conf/ijcai/GongZ16}, Gong et al. adopted attentional mechanism to scan input microblogs and select trigger words. Chen et al. \cite{DBLP:conf/sigir/ChenZ0NLC17} proposed item- and component-level attention mechanisms to model the implicit feedback in multimedia recommendation. In \cite{DBLP:conf/recsys/SeoHYL17}, Seo et al. proposed to model user preferences and item properties using convolutional neural networks (CNNs) with dual local and global attention, where the local attention provides insight on a user's preferences or an item's properties and the global attention helps CNNs focus on the semantic meaning of the review text.

However, our method is different from existing works. We propose a self-attentive encoder to further discriminate the user preference on checked-in POIs in multiple aspects. This is achieved by the proposed multi-dimensional attention mechanism, which utilizes an importance vector to depict the user preference. Furthermore, we adopt a neighbor-aware decoder to incorporate the geographical influence checked-in POIs applied on unvisited POIs, which makes the user reachability higher on the nearby neighbors of checked-in POIs. To the best of our knowledge, this is the first study to apply attention-based autoencoder in POI recommendation.

\section{Preliminaries}
In this section, we first introduce the definitions and notations. Then we review the basic ideas of autoencoders.

\subsection{Definition and Notation}
For ease of illustration, we first summarize the definitions and notations.

\textit{Definition 1}. (\textbf{POI}) A POI is defined as a uniquely identified site (e.g., a restaurant) with two attributes: an identifier and geographical coordinates (latitude and longitude).

\textit{Definition 2}. (\textbf{Check-in}) A check-in is a record that demonstrates a user has visited a POI at a certain time. A user's check-in is represented by a 3-tuple: user ID, POI ID, and the timestamp.

\textit{Definition 3}. (\textbf{POI Recommendation}) Given users' check-in records, POI recommendation aims at recommending a list of POIs for each user that the user is interested in but never visited.

\begin{table}[h]
\centering
\caption{List of notations.}
\label{tab:notations}
\begin{tabular}{ll}
 \hline
$ M $, $ N $ & the number of users and POIs \\
$ \mathbf{X} $, $ \hat{\mathbf{X}} $ & the input data and reconstructed data \\
$ \mathbf{R} $ & the check-in frequency matrix \\
$ \mathbf{C} $ & the confidence matrix \\
$ \mathbf{W}^{*} $, $ \mathbf{b}^{*} $ & the weight matrix and bias vector \\
$ a_{*} $ & the activation function \\
$ H $ & the dimension of the bottleneck layer \\
$ d_{a} $ & the dimension of the importance vector \\
$ \gamma $ & the parameter to control POI's correlation level \\
$ \alpha $, $ \epsilon $ & the parameters of the weighting scheme \\
$ \lambda $ & the regularization term \\ \hline
\end{tabular}
\end{table}

POI recommendation is commonly studied on a user-POI check-in matrix $ \mathbf{R} \in \mathbb{R}^{M \times N} $, where there are $ M $ users and $ N $ locations, each entry $ r_{u,i} $ represents the frequency user $ u $ checked-in at location $ i $. We denote the binary rating matrix as $ \mathbf{X} \in \mathbb{R}^{M \times N} $, where each entry $ x_{u,i} \in \{0,1\} $ indicates whether user $ u $ has visited location $ i $. The terms POI and location are used interchangeably in this paper. Here, following common symbolic notation, upper case bold letters denote matrices, lower case bold letters denote column vectors without any specification, and non-bold letters represent scalars. The notations are shown in Table \ref{tab:notations}.

\subsection{Autoencoders} \label{subsec:autoencoder}
A single hidden-layer autoencoder (AE) is an unsupervised neural network, which is composed of two parts, i.e., an encoder and a decoder. The encoder has one activation function that maps the input data to the latent space. The decoder also has one activation function mapping the representations in the latent space to the reconstruction space. Given the input $ \mathbf{x}_{i} $, a single hidden-layer autoencoder is shown as follows:
\begin{equation}
\begin{aligned}
& encoder: \mathbf{z}_{i} = a_{1}(\mathbf{W}^{(1)} \mathbf{x}_{i} + \mathbf{b}^{(1)}), \\
& decoder: \hat{\mathbf{x}}_{i} = a_{2}(\mathbf{W}^{(2)} \mathbf{z}_{i} + \mathbf{b}^{(2)}),
\end{aligned}
\label{eq:encoder_and_decoder}
\end{equation}
where $ \mathbf{W}^{*} $, $ \mathbf{b}^{*} $ and $ a_{*} $ denote the weight matrices, bias vectors and activation functions, respectively. $ \hat{\mathbf{x}}_{i} $ is the reconstructed version of $ \mathbf{x}_{i} $. The output $ \mathbf{z}_{i} $ of the encoder is the representation of $ \mathbf{x}_{i} $ in the latent space. The goal of the autoencoder is to minimize the reconstruction error of the output and the input. The loss function is shown as follows:
\begin{equation}
\mathcal{L}_{AE} = \sum_{i = 1}^{M}||\mathbf{x}_{i} - \hat{\mathbf{x}}_{i}||_{2}^{2}.
\label{eq:autoencoder_loss_function}
\end{equation}

\textbf{Relations to Matrix Factorization}. One reason that the autoencoder is capable of recommendation is that its formulation is much similar to the classical matrix factorization \cite{DBLP:journals/computer/KorenBV09}. Let we denote $ \mathbf{U} $ and $ \mathbf{V} $ as user latent factors and item latent factors, respectively, where $ \mathbf{V} $ has the same dimension with $ \mathbf{W}^{(2)}. $ If $ a_{2} $ is set to the identity function, then the formula in Eq. \ref{eq:encoder_and_decoder} can be rephrased as follows:
\begin{equation}
\begin{aligned}
 & \mathbf{u}_{i} = a_{1}(\mathbf{W}^{(1)} \mathbf{x}_{i} + \mathbf{b}^{(1)}), \\
 & \hat{\mathbf{r}}_{i} = \mathbf{V} \mathbf{u}_{i} + \mathbf{b}^{(2)} \Rightarrow \hat{r}_{i,j} = \mathbf{u}_{i}^{\top} \mathbf{v}_{j} + b_{j}^{(2)},
\end{aligned}
\label{eq:relation_to_MF}
\end{equation}
where $ \hat{r}_{i,j} $ is the predicted rating of a user $ i $ on an item $ j $, $ b_{j}^{(2)} $ is the $ j $-th element of $ \mathbf{b}^{(2)} $, which can be treated as the item bias. The rephrased formula demonstrates the strong relations between autoencoders and matrix factorization in recommendation, which makes autoencoders have the ability to recommend items.

\textbf{Relations to word2vec}. word2vec \cite{DBLP:conf/nips/MikolovSCCD13} is an effective and scalable method to learn embedding representations in word sequences, modeling words' contextual correlations in word sentences. word2vec utilizes either of two model architectures to produce a distributed representation of words: continuous bag-of-words (CBOW) or continuous skip-gram. Taking continuous skip-gram for example, the input of this model is an one-hot vector to represent the current word, then the model uses the current word to predict the surrounding window of context words. This model is highly similar to AE when the input of AE is an one-hot vector. If the current word is $ i $ and target word is $ j $, we set the activation function to identity and bias to zero, then the output of the decoder is:
\begin{equation}
\hat{x}_{ij} = \mathbf{W}_{j,*}^{(2) \top} \mathbf{W}_{*,i}^{(1)},
\label{eq:word2vec1}
\end{equation}
where $ \mathbf{W}_{*,i}^{(1)} $ and $ \mathbf{W}_{j,*}^{(2)} $ are the $ i $-th column and $ j $-th row of $ \mathbf{W}_{1} $ and $ \mathbf{W}_{2} $, respectively. If we further apply $ softmax $ on the output of the decoder:
\begin{equation}
P(l_{j}|l_{i}) = \frac{exp(\hat{x}_{ij})}{\sum_{j=1}^{N} exp(\hat{x}_{ij})},
\label{eq:word2vec2}
\end{equation}
where this probability shows how likely the word $ j $ will appear in the window of the current word $ i $. The combination of Eq. \ref{eq:word2vec1} and \ref{eq:word2vec2} is similar to the Eq. 2 in \cite{DBLP:conf/nips/MikolovSCCD13}. In our POI recommendation setting, this formula demonstrates if a user has checked-in location $ l_{i} $, how likely the user would check-in location $ l_{j} $. Therefore, the product of $ \mathbf{W}_{*,i}^{(1)} $ and $ \mathbf{W}_{j,*}^{(2)} $ can be used for capturing the relation between $ l_{i} $ and $ l_{j} $ in a single hidden-layer AE.

\section{Methodologies}
\begin{figure*}[t!]
    \centering
    \includegraphics[width=0.9\linewidth]{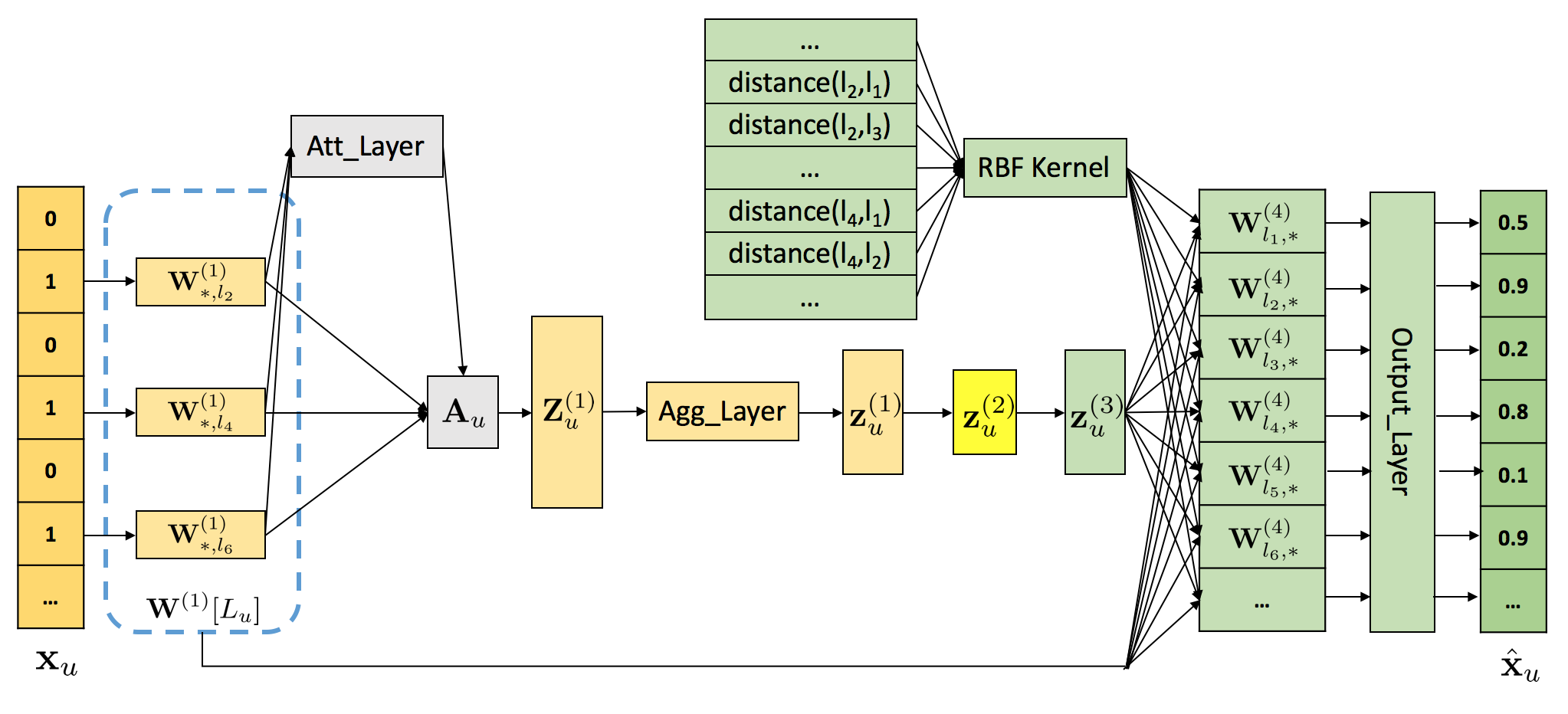}
    \caption{\label{fig:model_architecture}The model architecture. The yellow part is the self-attentive encoder, the green part is the neighbor-aware decoder, and the gray part is the attention network. The bright yellow rectangle is the user hidden representation. Specifically, \textit{Att\_Layer} denotes the attention layer and \textit{Agg\_Layer} denotes the aggregation layer.}
\vspace{-0.3cm}
\end{figure*}

In this section, we introduce the proposed model for POI recommendation, which consists of two components, i.e., a self-attentive encoder and a neighbor-aware decoder, demonstrating in Figure \ref{fig:model_architecture}. We first present the stacked autoencoder as our major building block. Then we illustrate the self-attentive encoder to adaptively select representative POIs that can reflect users' preferences. Next, we demonstrate the neighbor-aware decoder to model the geographical influence in POI recommendation, which is a phenomenon that users tend to check-in those unvisited POIs that close to a POI they checked-in before. Lastly, we present the loss function for implicit feedback and how to optimize the proposed model.

\subsection{Model Basics} \label{subsec:model_basics}
To learn the user hidden representation and reconstruct user preferences on unvisited POIs, we propose to adopt a stacked autoencoder, where the deep network architecture and non-linear activation functions can capture the non-linear and complex user-POI interactions \cite{DBLP:conf/www/HeLZNHC17}. Formally, the stacked autoencoder is shown as follows:
\begin{equation}
enc:
\begin{cases}
\mathbf{z}_{u}^{(1)} = a_{1}(\mathbf{W}^{(1)} \mathbf{x}_{u} + \mathbf{b}^{(1)}) \\
\mathbf{z}_{u}^{(2)} = a_{2}(\mathbf{W}^{(2)} \mathbf{z}_{u}^{(1)} + \mathbf{b}^{(2)})
\end{cases}
dec:
\begin{cases}
\mathbf{z}_{u}^{(3)} = a_{3}(\mathbf{W}^{(3)} \mathbf{z}_{u}^{(2)} + \mathbf{b}^{(3)}) \\
\hat{\mathbf{x}}_{u} = a_{4}(\mathbf{W}^{(4)} \mathbf{z}_{u}^{(3)} + \mathbf{b}^{(4)})
\end{cases}
\label{eq:deep_AE}
\end{equation}
where $ \mathbf{W}^{(1)} \in \mathbb{R}^{H_{1} \times N} $, $ \mathbf{W}^{(2)} \in \mathbb{R}^{H \times H_{1}} $, $ \mathbf{W}^{(3)} \in \mathbb{R}^{H_{1} \times H} $ and $ \mathbf{W}^{(4)} \in \mathbb{R}^{N \times H_{1}} $ are parameter matrices of the stacked AE. $ H_{1} $ is the dimension of the first hidden layer, and $ H $ is the dimension of the bottleneck layer. $ \mathbf{z}_{u}^{(2)} $ and $ \hat{\mathbf{x}}_{u} $ are the hidden representation and reconstructed ratings of user $ u $, respectively.

\subsection{Self-Attentive Encoder} \label{subsec:self-attention}
As presented in section \ref{subsec:model_basics}, we apply a stacked AE to learn users' hidden representations. In the proposed model, the input is a multi-hot user preference vector $ \mathbf{x}_{u} \in \mathbb{R}^{N} $, where 1 in the vector indicates the user has been to a certain POI. Based on the input, the encoder of a vanilla stacked AE works as follows: (1) given a user's check-in set $ L_{u} = \{l_{1},...,l_{n}\} $, where $ l_{n} $ is the index of a POI, corresponding POI vectors (e.g., $ \mathbf{W}_{*,l_n}^{(1)} $) in $ \mathbf{W}^{(1)} $ are selected and summed; (2) after having the summed vector, perform the activation function to get the user hidden representation. Here, $ \mathbf{W}^{(1)} $ works like a POI embedding matrix, which is similar to the word embedding matrix in the word2vec model.

Since the model input is a multi-hot vector, which makes each embedding in $ \mathbf{W}^{(1)}[L_{u}] $ equally contribute to the user hidden representation, where $ [\cdot] $ is the slicing operation that selects corresponding POI vectors to form a sub-matrix of $ \mathbf{W}^{(1)} $, which has the size $ H_{1} $-by-$ n $:
\begin{equation}
\mathbf{W}^{(1)}[L_{u}] = (\mathbf{W}_{*,l_1}^{(1)}, \mathbf{W}_{*,l_2}^{(1)},..., \mathbf{W}_{*,l_n}^{(1)}),
\label{eq:slice_operation}
\end{equation}
where $ \mathbf{W}_{*,l_n}^{(1)} $ is the $ l_n $-th column of $ \mathbf{W}^{(1)} $.

However, in the user check-in history, there should be some POIs more representative than others that can directly reflect a user's preferences. These representative POIs should contribute more to the user hidden representation to express the user preference. This inspires us to propose a self-attentive mechanism, which learns a weighted sum of embeddings in $ \mathbf{W}^{(1)}[L_{u}] $ to form a user's hidden representation.

The goal of the self-attentive encoder is to adaptively assign different importances on checked-in POIs for expressing various users' preference levels. Then the embeddings of checked-in POIs are aggregated in a weighted manner to characterize users. Given checked-in POI embeddings $ \mathbf{W}^{(1)}[L_{u}] $ of user $ u $, we use a single-layer network without bias to compute the importance score (attention score):
\begin{equation}
\mathbf{a}_{u} = softmax(tanh(\mathbf{w}_{a}^{\top} \mathbf{W}^{(1)}[L_{u}]),
\label{eq:attention_vector}
\end{equation}
where $ \mathbf{w}_{a} \in \mathbb{R}^{H_{1}} $ is the parameter in the attention layer, the $ softmax $ ensures all the computed weights sum up to 1. Then we sum up the embeddings in $ \mathbf{W}^{(1)}[L_{u}] $ according to the importance score provided by $ \mathbf{a}_{u} $ to get a vector representation of the user:
\begin{equation}
\mathbf{z}_{u}^{(1)} = \sum_{l_j \in L_u} a_{u,j} \mathbf{W}_{*,l_j}^{(1)}.
\end{equation}

However, the standard attention mechanism that assigning a single importance value to a POI makes the model focus on only one specific aspect of POIs \cite{DBLP:journals/corr/LinFSYXZB17}, which is not sufficient to reflect the sophisticated human sentiment on POIs. Taking a restaurant for example. From the perspective of food flavor, a user likes this restaurant; from the perspective of eating environment, the user may think the restaurant is not good enough. Thus, to capture the user preference from different aspects, we may need to perform multiple times of Eq. \ref{eq:attention_vector} with different sets of parameters.

Therefore, we adopt an importance score matrix to capture the effects of multiple-dimensional attention \cite{DBLP:conf/nips/VaswaniSPUJGKP17} on POIs. Each dimension of the importance scores represents the importance levels of checked-in POIs in a certain aspect. Suppose we want $ d_{a} $ aspects of attention to be extracted from the embeddings, then we can extend $ \mathbf{w}_{a} $ to $ \mathbf{W}_{a} \in \mathbb{R}^{d_{a} \times H_{1} } $\footnote{We also tried using a two-layer neural network to compute the importance score matrix, which achieves similar performance with the single-layer one.}:
\begin{equation}
\mathbf{A}_{u} = softmax(tanh(\mathbf{W}_{a} \mathbf{W}^{(1)}[L_{u}]),
\label{eq:attention_matrix}
\end{equation}
where $ \mathbf{A}_{u} \in \mathbb{R}^{d_{a} \times n} $ is the importance score matrix, each column of $ \mathbf{A}_{u} $ is the importance vector of a specific POI, and each row of $ \mathbf{A}_{u} $ depicts the importance levels of $ n $ checked-in POIs in a certain aspect. The $ softmax $ is performed along the second dimension of its input. By multiplying the importance score matrix with the POI embeddings, we have:
\begin{equation}
\mathbf{Z}_{u}^{(1)} = \mathbf{A}_{u} \cdot (\mathbf{W}^{(1)}[L_{u}])^{\top},
\label{eq:user_embedding_matrix}
\end{equation}
where $ \mathbf{Z}_{u}^{(1)} \in \mathbb{R}^{d_{a} \times H_{1}} $ is the matrix representation of user $ u $, which depicts the user from $ d_{a} $ aspects. To make the matrix representation of users fit our encoder, we have one more neural layer to aggregate users' representations from different aspects into one aspect. Then the vector representation of user $ u $ is shown:
\begin{equation}
\mathbf{z}_{u}^{(1)} = a_{t}(\mathbf{Z}_{u}^{(1) \top} \mathbf{w}_{t} + \mathbf{b}_{t}),
\label{eq:user_embedding}
\end{equation}
where $ \mathbf{w}_{t} \in \mathbb{R}^{d_{a}} $ is the parameter in the aggregation layer.

\subsection{Neighbor-Aware Decoder} \label{subsec:neighbor_influence}
In LBSNs, there is physical distance between users and POIs, which is a unique property distinguishing POI recommendation from other recommendation tasks. In a user's check-in history, the user's occurrences are typically constrained in several certain areas. This is the well-known geographical clustering phenomenon (a.k.a geographical influence) in users' check-in activities, which has been exploited to largely improve the POI recommendation performance \cite{DBLP:conf/sigir/YeYLL11,DBLP:conf/aaai/ChengYKL12,DBLP:conf/cikm/LiuWSM14,DBLP:conf/kdd/LianZXSCR14,DBLP:conf/sigir/LiCLPK15,DBLP:conf/kdd/LiGHZ16}. Different from most of the previous studies that mainly exploit geographical influence from a user's perspective: learning the geographical distribution of each individual user's check-ins \cite{DBLP:conf/sigir/YeYLL11,DBLP:conf/aaai/ChengYKL12,DBLP:conf/kdd/LiGHZ16} or modeling the user preference on a POI from both this POI and its neighbors \cite{DBLP:conf/kdd/LianZXSCR14,DBLP:conf/sigir/LiCLPK15}, the proposed neighbor-aware influence model captures the geographical influence solely from the perspective of POIs. 

According to aforementioned geographical influence, one intuition contributes to this phenomenon: users prefer to check-in POIs surrounded a POI that they visited before. From this intuition, a checked-in POI may have impacts on other unvisited POIs, and the impact level is determined by the properties and distance of two POIs. Inspired by the skip-gram model of word2vec, which applies the inner product to predict the context words given an input word, we also leverage similar techniques to model the influence a checked-in POI exerted on unvisited POIs (section \ref{subsec:autoencoder}, relations to word2vec). The proposed technique can discover unvisited POIs that may be similar and close to the visited POIs. Similarly, we treat $ \mathbf{W}^{(1)} $ as the POI embedding matrix (the first weight matrix in word2vec) and $ \mathbf{W}^{(4)} $ as the context POI embedding matrix (the second weight matrix in word2vec). Moreover, the proposed method is also similar to FISM \cite{DBLP:conf/kdd/KabburNK13}, where FISM adopts two matrices of item latent factors to model the similarity between items.

Formally, given a user's check-in set $ L_{u} = \{l_{1},...,l_{n}\} $, the influence checked-in POIs exerted on unvisited POIs is shown:
\begin{equation}
\mathbf{P}_{u} = \mathbf{W}^{(4)} \cdot \mathbf{W}^{(1)}[L_{u}],
\end{equation}
where $ \mathbf{P}_{u} \in \mathbb{R}^{N \times n} $. Each column of $ \mathbf{P}_{u} $ is the influence a certain checked-in POI applied on all other POIs (the influence on itself is set to 0).

The above inner product gives a basic indication about how related two POIs are, however, it does not explicitly take the distance between two POIs into consideration. According to Tobler's First Law of Geography, everything is related to everything else, but near things are more related than distant things. To incorporate the geographical distance property, we adopt the Gaussian radial basis function kernel (RBF kernel) to further make checked-in POIs exert more influence on nearby unvisited POIs. The RBF kernel is shown as follows:
\begin{equation}
K(l_{i},l_{j}) = exp(-\gamma ||\mathbf{l}_{i} - \mathbf{l}_{j}||^{2}),
\label{eq:rbf_kernel}
\end{equation}
where $ \mathbf{l}_{i} $ and $ \mathbf{l}_{j} $ are the geographical coordinates of two POIs $ l_{i} $ and $ l_{j} $. $ \gamma > 0 $ is a hyper-parameter to control the geographical correlation level of two given POIs, a larger value of $ \gamma $ will lead to a larger $ K(l_{i},l_{j}) $. The value range of RBF kernel is $ K(l_{i},l_{j}) \in [0, 1] $. For computation simplicity, if the value of $ K(l_{i},l_{j}) $ is less than 0.1, we set it to 0. We can pre-compute the pair-wise RBF value of each POI pair to get a RBF value matrix $ \mathbf{K} \in \mathbb{R}^{N \times N} $.
% This setting makes the influence radius of a checked-in POI to be 15km. 

By incorporating the RBF kernel, our neighbor-aware influence model is shown:
\begin{equation}
\mathbf{P}_{u} = (\mathbf{W}^{(4)} \cdot \mathbf{W}^{(1)}[L_{u}]) \odot \mathbf{K}[L_{u}],
\label{eq:neighbor_influence_matrix}
\end{equation}
where $ \mathbf{K}[L_{u}] \in \mathbb{R}^{N \times n} $ is the RBF kernel value from Eq. \ref{eq:rbf_kernel}, $ \odot $ is the element-wise multiplication. 

To obtain the accumulated influence from all checked-in POIs, we sum along the row of $ \mathbf{P}_{u} \in \mathbb{R}^{N \times n} $ to get $ \mathbf{p}_{u} \in \mathbb{R}^{N} $:
\begin{equation}
\mathbf{p}_{u} = \sum_{j=1}^{n} \mathbf{P}_{u} ^ {(i,j)}, i=1,2,...,N,
\label{eq:neighbor_influence_vector}
\end{equation}
where $ i $ and $ j $ are the row and column index, respectively.

To incorporate the neighbor-aware influence, the decoder of the proposed model can be rewritten as:
\begin{equation}
\hat{\mathbf{x}}_{u} = a_{4}(\mathbf{W}^{(4)} \mathbf{z}_{u}^{(3)} + \mathbf{p}_{u} + \mathbf{b}^{(4)}),
\label{eq:neighbor_aware_decoder}
\end{equation}
where $ \mathbf{W}^{(4)} \mathbf{z}_{u}^{(3)} $ captures the user preference, $ \mathbf{p}_{u} $ models the neighbor-aware geographical influence.

\textbf{Discussion}. As we mentioned before, the way we adopt the inner product to capture the relations between POIs is similar to FISM \cite{DBLP:conf/kdd/KabburNK13}, if we treat $ \mathbf{W}^{(1)} $ as $ \mathbf{P} $ and $ \mathbf{W}^{(4)} $ as $ \mathbf{Q} $ in FISM. In FISM, the predicted rating of user $ u $ on item $ i $ is mainly estimated by $ \sum_{j \in \mathcal{R}_{u}^{+}} \mathbf{p}_{j} \mathbf{q}_{i}^{\top} $, where $ \mathcal{R}_{u}^{+} $ is the set of items rated by user $ u $, $ \mathbf{p}_{j} $ and $ \mathbf{q}_{i} $ are learned item latent factors from $ \mathbf{P} $ and $ \mathbf{Q} $, respectively. % Otherwise, our method is also similar to the alignment function of attention mechanism in machine translation. In machine translation, researchers apply an alignment model and $ softmax $ \cite{DBLP:journals/corr/BahdanauCB14,DBLP:conf/emnlp/LuongPM15} to compute an attention weight vector between the target word's hidden state and the source words' hidden states. If we treat $ \mathbf{W}^{(1)} $ as the source hidden states and $ \mathbf{W}^{(4)} $ as the target hidden states, each row of $ (\mathbf{W}^{(4)} \cdot \mathbf{W}^{(1)}[L_{u}]) \in \mathbb{R}^{N \times n} $ may behave like an attention weight vector. The attention weight vector can tell if a user wants to check-in an unvisited POI, which POI the user has checked-in plays an important role.

\subsection{Weighted Loss for Implicit Feedback} \label{subsec:weighted_AE}
% Generally, there are two kinds of data used to capture user preferences in recommendation tasks, i.e., explicit feedback and implicit feedback. Explicit feedback directly reflects how much a user prefers an item, such as star ratings on Netflix. In such cases, users not only reveal their favorite movies with high ratings, but also their least favorite movies with low ratings. 

In POI recommendation, check-in data is treated as implicit feedback. Since a user's check-in records only include the locations she visited, and the visit frequency indicates the confidence level of her preference. Therefore, there are only \textit{positive} examples observed in the check-in records, which makes POI recommendation an One Class Collaborative Filtering (OCCF) problem \cite{DBLP:conf/icdm/PanZCLLSY08,DBLP:conf/icdm/HuKV08}. 

To tackle the OCCF problem and capture user preferences from check-in data, we adopt a general weighting scheme \cite{DBLP:conf/icdm/HuKV08} to distinguish visited and unvisited POIs. Specifically, we consider all unvisited locations as negative examples, and assign the weights of all negative examples to the same value, e.g., 1. As for visited locations, the weights are increased monotonically with users' check-in frequencies. With such a weighting scheme, our model not only distinguishes visited and unvisited POIs, but also discriminates the confidence levels of all visited POIs. The objective function for implicit feedback is presented as follows,
\begin{equation}
\mathcal{L}_{WAE} = \sum_{u = 1}^{M} \sum_{i=1}^{N}||c_{u,i} (x_{u,i} - \hat{x}_{u,i})||_{2}^{2} = ||\mathbf{C} \odot (\mathbf{X}-\mathbf{\hat{X}})||^{2}_{F},
\label{eq:weighted_autoencoder_loss_function}
\end{equation}
where $ \odot $ is the element-wise multiplication of matrices. $ ||\cdot||_{F} $ is the Frobenius norm of matrices. In particular, we set the confidence matrix $ \mathbf{C} \in \mathbb{R}^{M \times N} $ as follows:
\begin{equation}
c_{u,i}=
\begin{cases}
1 + \alpha log (1+r_{u,i} / \epsilon ) & \text{ if } r_{u,i} > 0 \\ 
1 & \text{} otherwise
\end{cases}
\label{eq:weight_function}
\end{equation}
where $ \alpha $ and $ \epsilon $ are hyper-parameters. This setting exactly encodes the observation that the frequency is a confidence of user preferences. This weighted loss with a vanilla autoencoder can be used in other recommendation tasks that take implicit feedback as input.

\subsection{Network Training}
By combining regularization terms, the objective function of the proposed model is shown as follows:
\begin{equation}
\begin{aligned}
\mathcal{L} =||\mathbf{C} \odot (\mathbf{X}-\mathbf{\hat{X}})||^{2}_{F} + \lambda(||\mathbf{W}^{*}||^{2}_{F} + ||\mathbf{W}_{a}||^{2}_{F} + ||\mathbf{w}_{t}||^{2}_{2}),
\end{aligned}
\label{eq:final_loss}
\end{equation}
where $ \lambda $ is the regularization parameter, $ \mathbf{W}^{*} $ includes $ \mathbf{W}^{(1)} $, $ \mathbf{W}^{(2)} $, $ \mathbf{W}^{(3)} $ and $ \mathbf{W}^{(4)} $. $ \mathbf{W}_{a} $ and $ \mathbf{w}_{t} $ are the learned parameters in the attention layer and aggregation layer, respectively. By minimizing the objective function, the partial derivatives with respect to all the parameters can be computed by gradient descent with back-propagation. And we apply Adam \cite{DBLP:journals/corr/KingmaB14} to automatically adapt the learning rate during the learning procedure. The mini-batch training algorithm is shown in Alg. \ref{alg:batch_training_algorithm}. 

\begin{algorithm}
  \textbf{Input}: $ \mathbf{X} $, $ \mathbf{C} $ \;
  % \KwIn{$ \mathbf{X} $, $ \mathbf{C} $, $ I $}
  % \KwOut{the reconstructed matrix $ \hat{\mathbf{X}} $}
  Initialize parameters $ \mathbf{W}^{*} $, $ \mathbf{W}_{a} $, $ \mathbf{w}_{t} $, $ \mathbf{b}^{*} $, $ \mathbf{b}_{t} $ \;
  numBatches = $ M $ $/$ $ batchSize $ \;
  \While{iter < numIterations} {
    Shuffle($ \mathbf{X}, \mathbf{C} $) \;
    \For{batchID = 0; batchID < numBatches; batchID++} {
      $ \mathbf{X}_{batch}, \mathbf{C}_{batch} $ = ExtractBatchData(batchID, $ \mathbf{X}, \mathbf{C} $) \;
      Apply Eq. \ref{eq:slice_operation} to get $ \mathbf{W}^{(1)}[L_{u}] $ for each user $ u $ in $ \mathbf{X}_{batch} $ \;
      Apply Eq. \ref{eq:attention_matrix}, Eq. \ref{eq:user_embedding_matrix} and Eq. \ref{eq:user_embedding} to get $ \mathbf{z}_{u}^{(1)} $ \;
      Apply Eq. \ref{eq:neighbor_influence_matrix} and Eq. \ref{eq:neighbor_influence_vector} to get $ \mathbf{p}_{u} $ \;
      %Concatenate each $ \mathbf{z}_{u}^{(1)} $ and $ \mathbf{p}_{u} $ in the batch, respectively \;
      Apply Eq. \ref{eq:deep_AE} and Eq. \ref{eq:neighbor_aware_decoder} to get $ \hat{\mathbf{X}}_{batch} $ \;
      Apply Eq. \ref{eq:final_loss} to obtain $ \mathcal{L}_{batch} $ and back-propagate the error through the entire network \;
    }
  }
  % \Return{$ \hat{\mathbf{X}} $}

\caption{Training Algorithm}
\label{alg:batch_training_algorithm}
\end{algorithm}

\textbf{Recommendation}. At prediction time, the proposed model takes each user's binary rating vector $ \mathbf{x}_{u} $ as input and obtains the reconstructed rating vector $ \hat{\mathbf{x}}_{u} $ as output. Then the POIs that are not in training set and have largest prediction scores in $ \hat{\mathbf{x}}_{u} $ are recommended to the user.

\section{Experiments}
In this section, we evaluate the proposed model with the state-of-the-art methods on three real-world datasets.

\subsection{Datasets}
We evaluate the proposed model on three real-world datasets: Gowalla \cite{DBLP:conf/kdd/ChoML11}, Foursquare \cite{DBLP:journals/pvldb/LiuPCY17} and Yelp \cite{DBLP:journals/pvldb/LiuPCY17}. The Gowalla dataset was generated worldwide from February 2009 to October 2010. The Foursquare dataset comprises check-ins from April 2012 to September 2013 within the United States (except Alaska and Hawaii). The Yelp dataset is obtained from the Yelp dataset challenge round 7. Each check-in record in above datasets includes a timestamp, a user ID, a POI ID, and the latitude and longitude of this POI. 

To filter noisy data, for the Gowalla dataset, we remove users whose total check-ins are less than 20 and POIs visited less than 20 times; for the Foursquare and Yelp datasets, we eliminate those users with fewer than 10 check-in POIs, as well as those POIs with fewer than 10 visitors. The data statistics after preprocessing are shown in Table \ref{tab:data_statistics}. For each user, we randomly select 20\% of her visiting locations as ground truth for testing. The remaining constitutes the training set. Similar data partition methods have been widely used in previous work \cite{DBLP:conf/kdd/LianZXSCR14,DBLP:conf/aaai/GaoTHL15,DBLP:conf/sigir/YeYLL11} to validate the performance of POI recommendation. The random selection is carried out six times independently, we tune the model on one partition and report the average results on the rest five partitions.

\begin{table}[ht]
\centering
\caption{\label{tab:data_statistics}The statistics of datasets.}
\begin{tabular}{ |c|c|c|c|c|c|c| }
 \hline
 Dataset & \#Users & \#POIs & \#Check-ins & Density \\
 \hline
 Gowalla & 43,074 & 46,234 & 1,720,082 & 0.0500\% \\ 
 \hline
 Foursquare & 24,941 & 28,593 & 1,196,248 & 0.1006\% \\ 
 \hline
 Yelp & 30,887 & 18,995 & 860,888 & 0.1399\% \\
 \hline
\end{tabular}
\vspace{-0.3cm}
\end{table}

\subsection{Evaluation Metrics}
We evaluate our model versus other models in terms of Precision@k, Recall@k and MAP@k. For each user, Precision@k indicates what percentage of locations among the top $ k $ recommended POIs has been visited by her, while Recall@k indicates what percentage of her visiting locations can emerge in the top $ k $ recommended POIs. MAP@k is the mean average precision at $ k $, where average precision is the average of precision values at all ranks where relevant POIs are found. They are formally defined as follows,
\begin{comment}
\begin{equation}
Precision@k=\frac{1}{M} \sum_{i=1}^{M} \frac{S_{i}(k) \cap T_{i}}{k},
\label{eq:precision_at_k}
\end{equation}

\begin{equation}
Recall@k=\frac{1}{M} \sum_{i=1}^{M} \frac{S_{i}(k) \cap T_{i}}{|T_{i}|},
\label{eq:recall_at_k}
\end{equation}

\begin{equation}
MAP@k=\frac{1}{M} \sum^{M}_{i=1} \frac{\sum^{k}_{j=1} p(j) \times rel(j)}{|T_{i}|},
\label{eq:map_at_k}
\end{equation}
\end{comment}
\begin{equation}
\begin{aligned}
Precison@k & =\frac{1}{M} \sum_{i=1}^{M} \frac{S_{i}(k) \cap T_{i}}{k},
Recall@k =\frac{1}{M} \sum_{i=1}^{M} \frac{S_{i}(k) \cap T_{i}}{|T_{i}|}, \\
MAP@k & =\frac{1}{M} \sum^{M}_{i=1} \frac{\sum^{k}_{j=1} p(j) \times rel(j)}{|T_{i}|},
\end{aligned}
\end{equation}
where $ S_{i}(k) $ is a set of top-$ k $ unvisited locations recommended to user $ i $ excluding those locations in the training, and $ T_{i} $ is a set of locations that are visited by user $ i $ in the testing. $ p(j) $ is the precision of a cut-off rank list from $ 1 $ to $ j $, and $ rel(j) $ is an indicator function that equals to $ 1 $ if the location is visited in the testing, otherwise equals to $ 0 $.

\subsection{Methods Studied}
To demonstrate the effectiveness of our model, we compare to the following POI recommendation methods.

\textit{Traditional MF methods for implicit feedback}\footnote{The implementations are from LibRec: https://www.librec.net/}:
\begin{itemize}
\item \textbf{WRMF}, weighted regularized matrix factorization \cite{DBLP:conf/icdm/HuKV08}, which minimizes the square error loss by assigning both observed and unobserved check-ins with different confidential values based on matrix factorization.
\item \textbf{BPRMF}, Bayesian personalized ranking \cite{DBLP:conf/uai/RendleFGS09}, which optimizes the ordering of the preferences for the observed locations and the unobserved locations.
\end{itemize}

\textit{Classical POI recommendation methods\footnote{In a recent study \cite{DBLP:journals/pvldb/LiuPCY17} that evaluated a number of POI recommendation methods, RankGeoFM and IRENMF achieve the best results on three datasets.}}:
\begin{itemize}
\item \textbf{MGMMF}, a multi-center Gaussian model fused with matrix factorization \cite{DBLP:conf/aaai/ChengYKL12}, which learns regions of activities for each user using multiple Gaussian distributions.
\item \textbf{IRENMF}, instance-region neighborhood matrix factorization \cite{DBLP:conf/cikm/LiuWSM14}, which incorporates instance-level and region-level geographical influence into weighted matrix factorization.
\item \textbf{RankGeoFM}, ranking-based geographical factorization \cite{DBLP:conf/sigir/LiCLPK15}, which is an ranking-based matrix factorization model that learns users' preference rankings for POIs, as well as includes the geographical influence of neighboring POIs.
\end{itemize}

\textit{Deep learning-based methods}:
\begin{itemize}
\item \textbf{PACE}, preference and context embedding \cite{DBLP:conf/kdd/YangBZY017}, a deep neural architecture that jointly learns the embeddings of users and POIs to predict both user preference over POIs and various context associated with users and POIs.
\item \textbf{DeepAE}, a three-hidden-layer autoencoder with a weighted loss function (section \ref{subsec:weighted_AE}).
\end{itemize}

\textit{The proposed method}:
\begin{itemize}
\item \textbf{SAE-NAD}, the proposed model with self-attentive encoder (section \ref{subsec:self-attention}) and neighbor-aware decoder (section \ref{subsec:neighbor_influence}) for implicit feedback (section \ref{subsec:weighted_AE}).
\end{itemize}

\begin{figure*}[t!]
    \centering
    \begin{subfigure}[t]{0.33\textwidth}
        \centering
        \includegraphics[width=\linewidth]{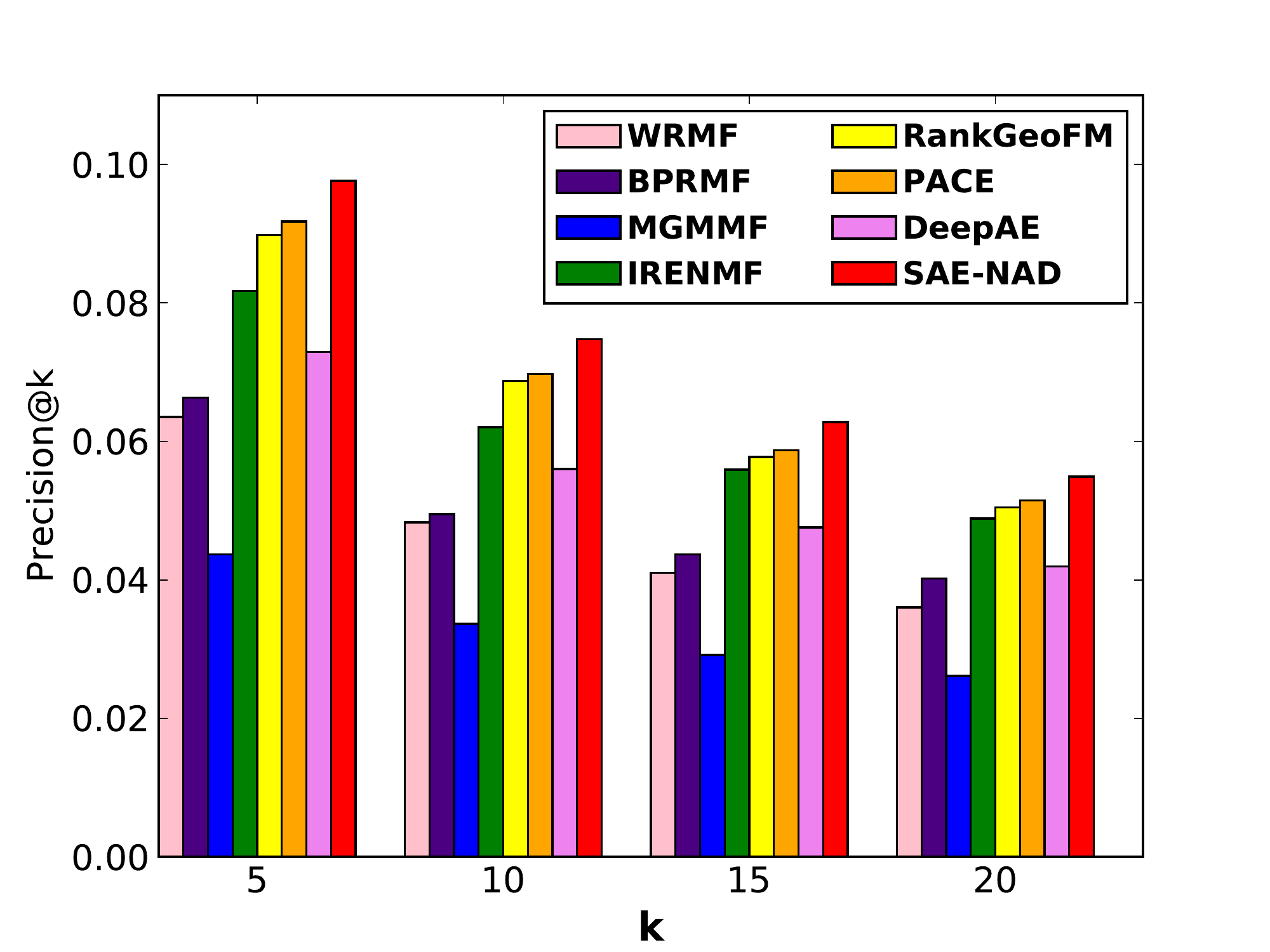}
        \caption{\label{fig:precision_Gowalla}Precision@k on Gowalla}
    \end{subfigure}%
    \begin{subfigure}[t]{0.33\textwidth}
        \centering
        \includegraphics[width=\linewidth]{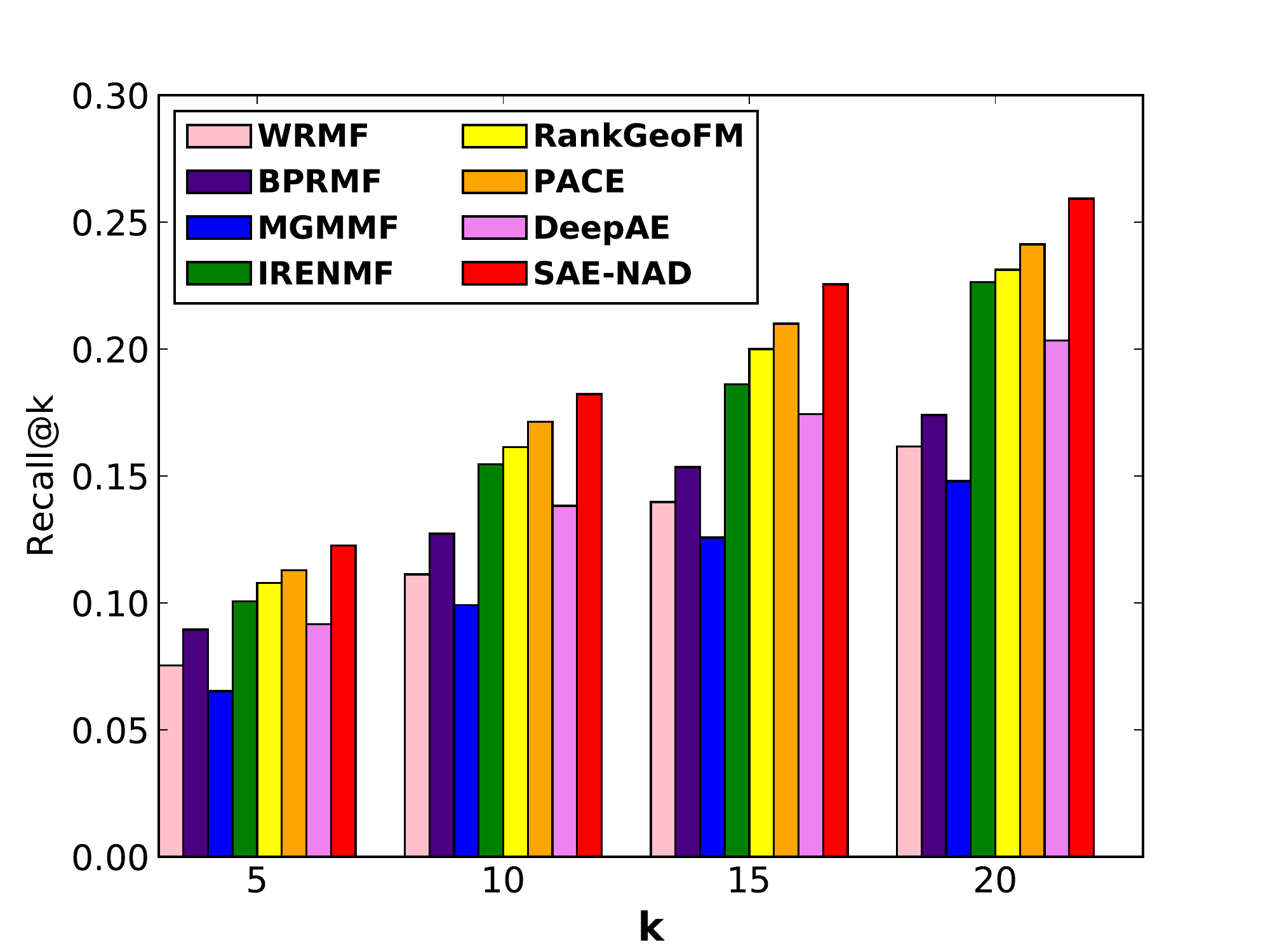}
        \caption{\label{fig:recall_Gowalla}Recall@k on Gowalla}
    \end{subfigure}%
    \begin{subfigure}[t]{0.33\textwidth}
        \centering
        \includegraphics[width=\linewidth]{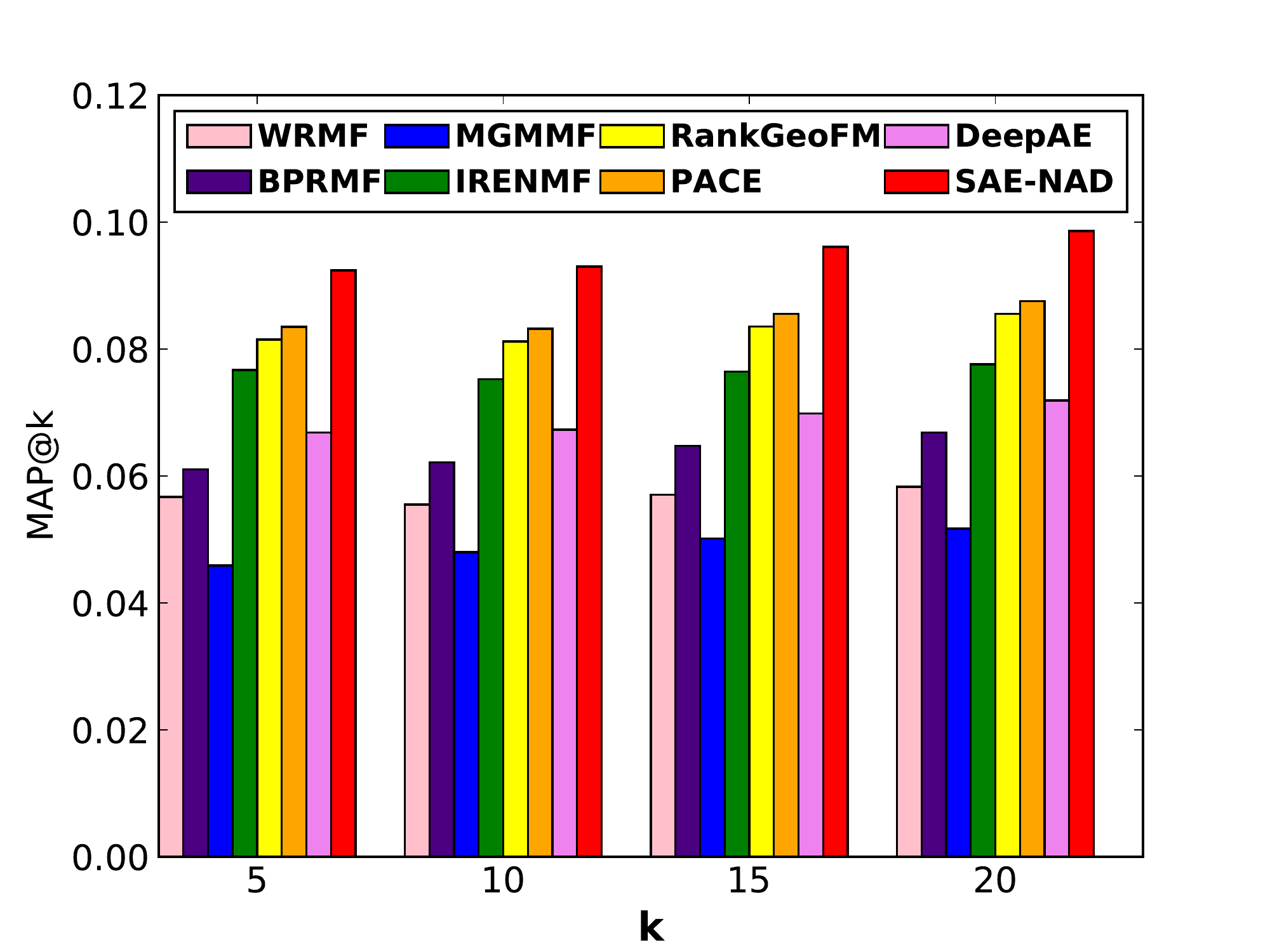}
        \caption{\label{fig:map_Gowalla}MAP@k on Gowalla}
    \end{subfigure}
    \caption{\label{fig:comparison_Gowalla}The comparison of performance on Gowalla.}
\vspace{-0.3cm}
\end{figure*}

\begin{figure*}[t!]
    \centering
    \begin{subfigure}[t]{0.33\textwidth}
        \centering
        \includegraphics[width=\linewidth]{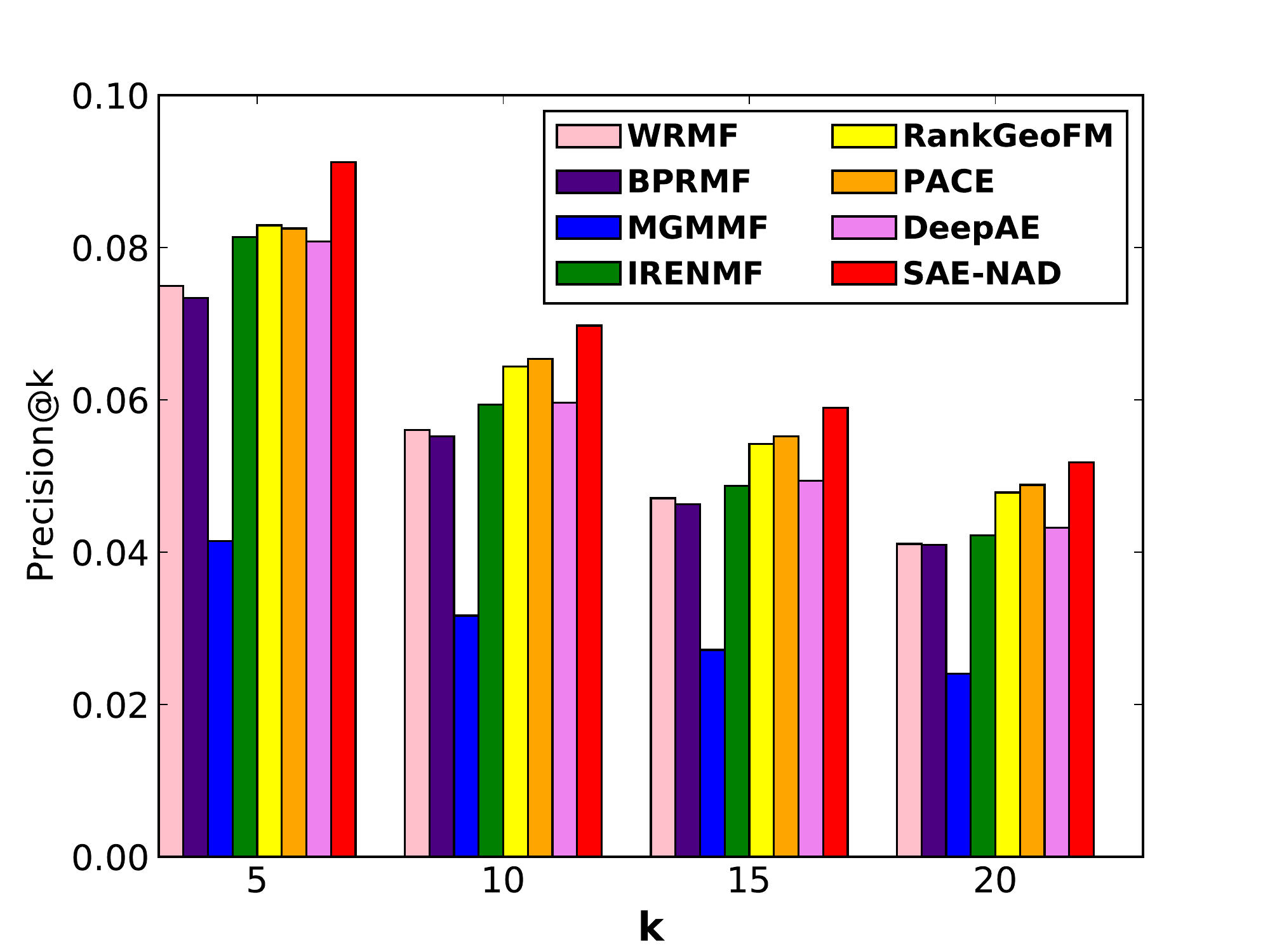}
        \caption{\label{fig:precision_Foursquare}Precision@k on Foursquare}
    \end{subfigure}%
    \begin{subfigure}[t]{0.33\textwidth}
        \centering
        \includegraphics[width=\linewidth]{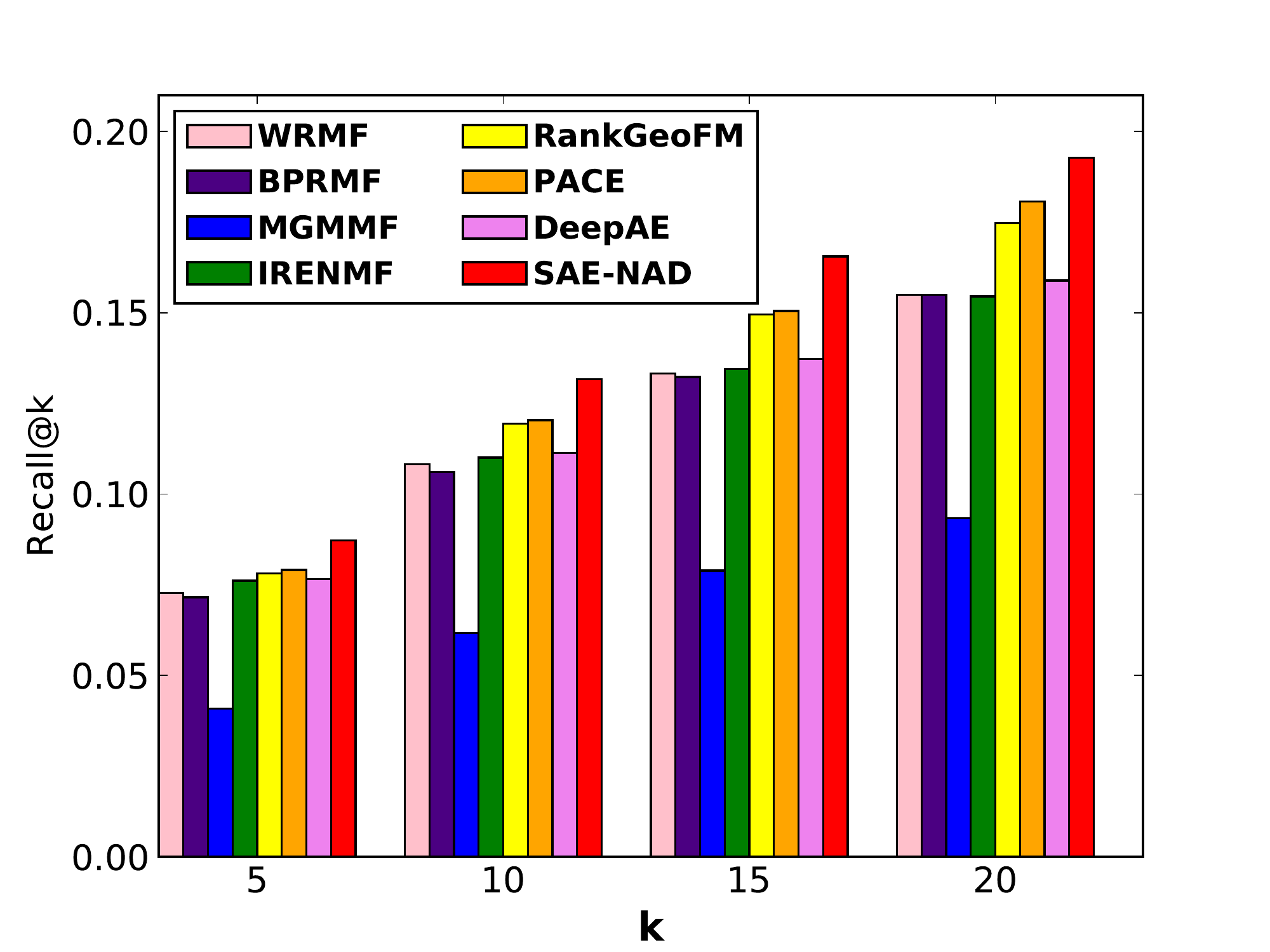}
        \caption{\label{fig:recall_Foursquare}Recall@k on Foursquare}
    \end{subfigure}%
    \begin{subfigure}[t]{0.33\textwidth}
        \centering
        \includegraphics[width=\linewidth]{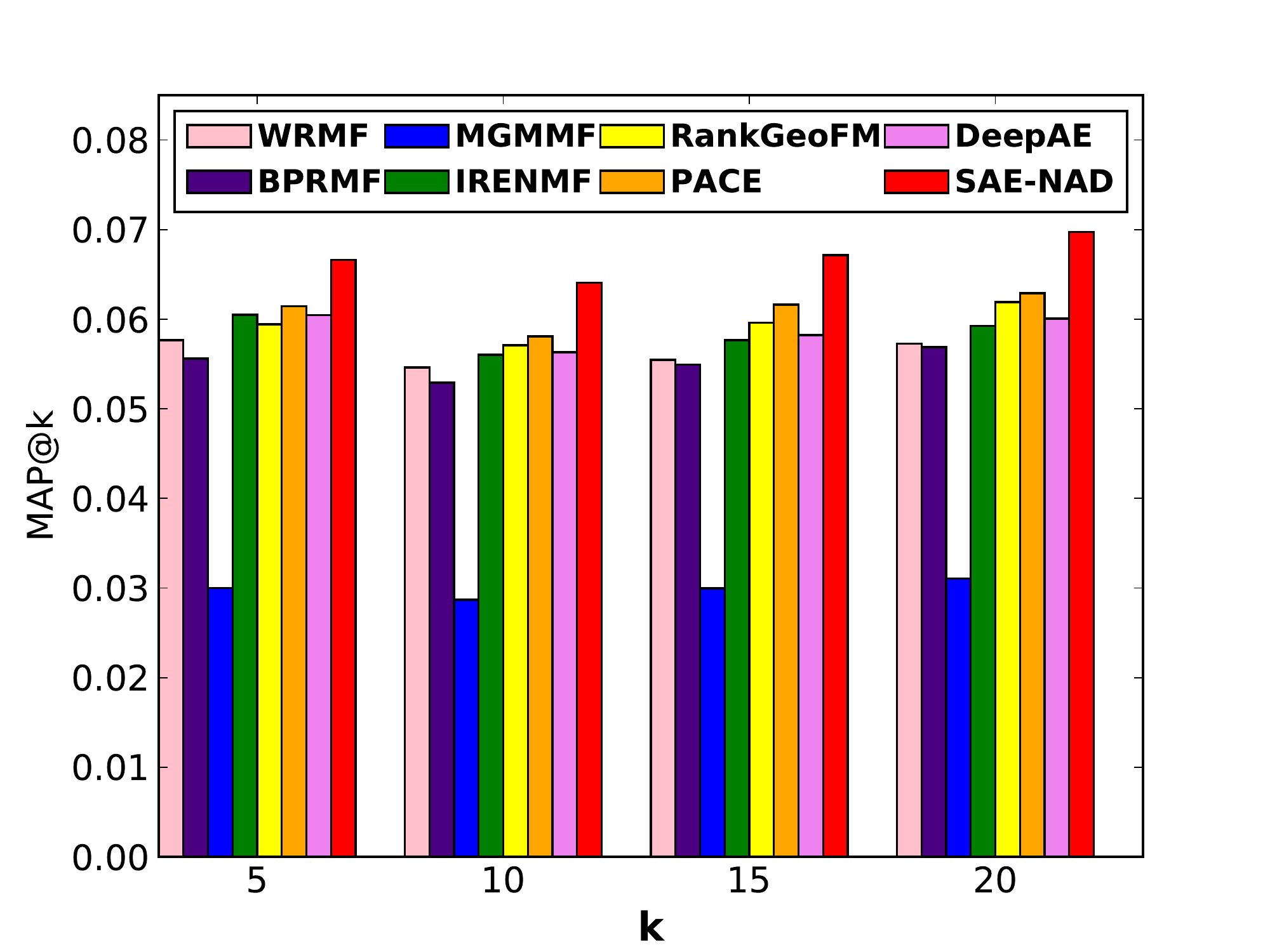}
        \caption{\label{fig:map_Foursquare}MAP@k on Foursquare}
    \end{subfigure}
    \caption{\label{fig:comparison_Foursquare}The comparison of performance on Foursquare.}
\vspace{-0.3cm}
\end{figure*}

\begin{figure*}[t!]
    \centering
    \begin{subfigure}[t]{0.33\textwidth}
        \centering
        \includegraphics[width=\linewidth]{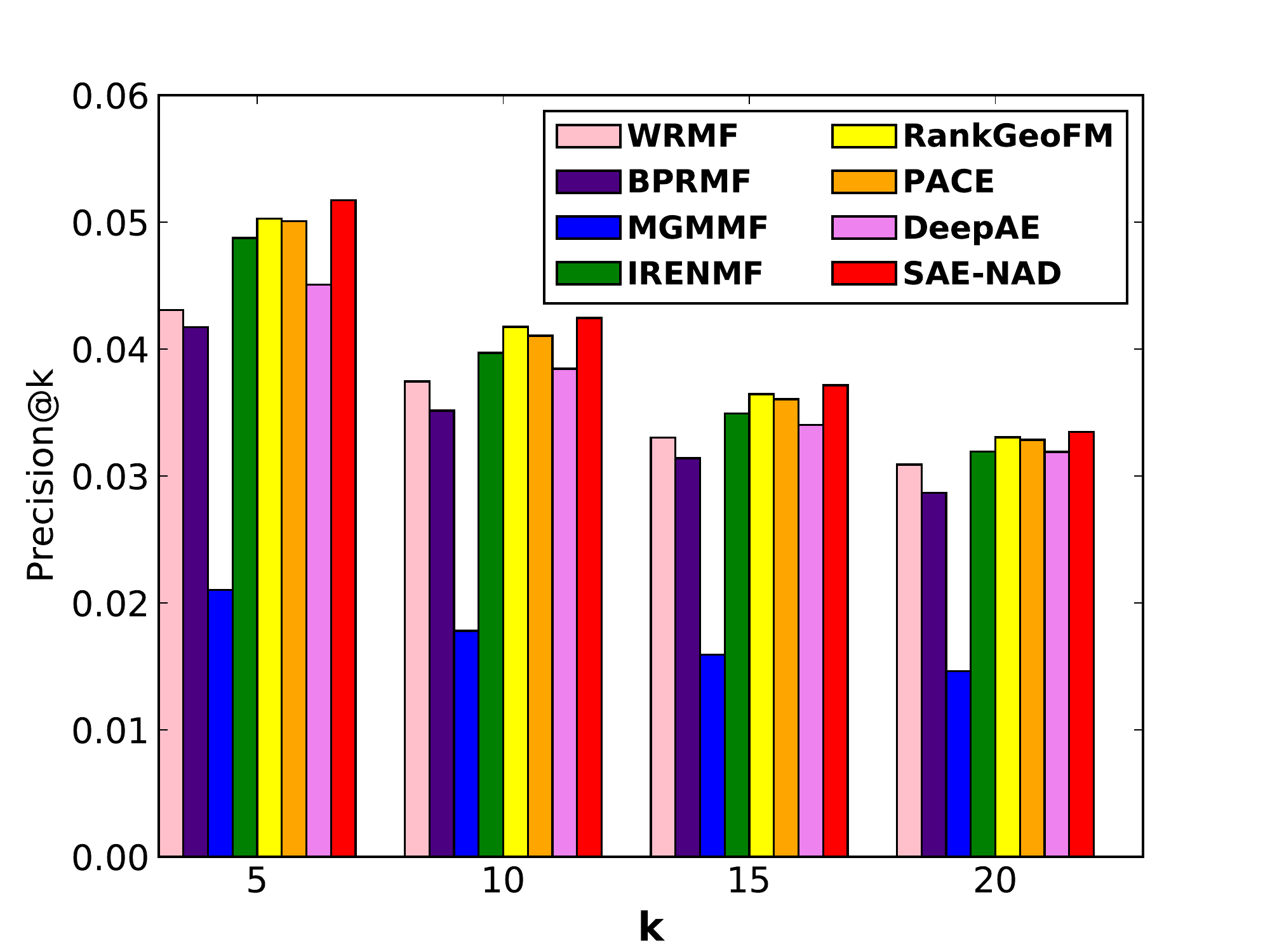}
        \caption{\label{fig:precision_Yelp}Precision@k on Yelp}
    \end{subfigure}%
    \begin{subfigure}[t]{0.33\textwidth}
        \centering
        \includegraphics[width=\linewidth]{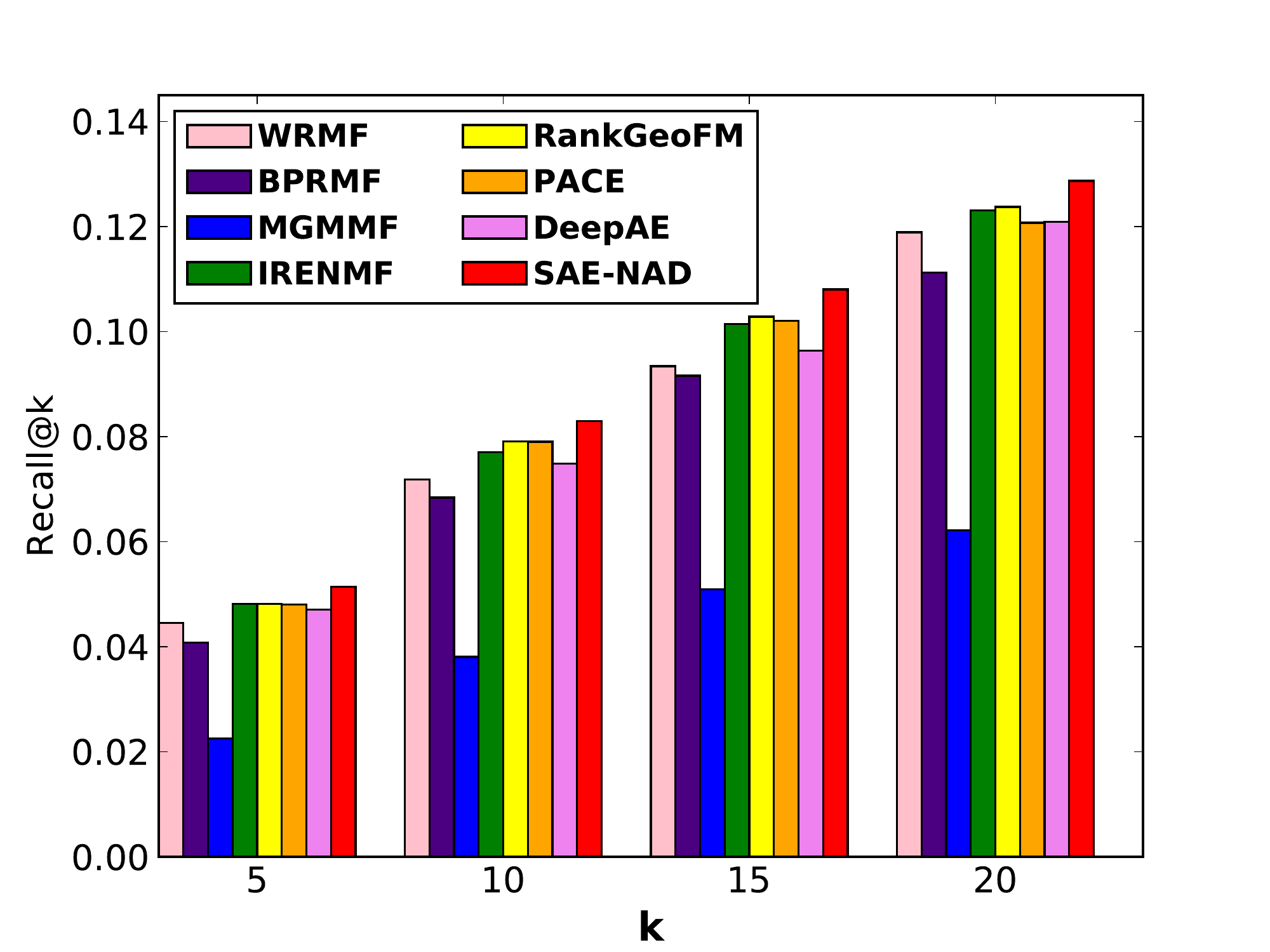}
        \caption{\label{fig:recall_Yelp}Recall@k on Yelp}
    \end{subfigure}%
    \begin{subfigure}[t]{0.33\textwidth}
        \centering
        \includegraphics[width=\linewidth]{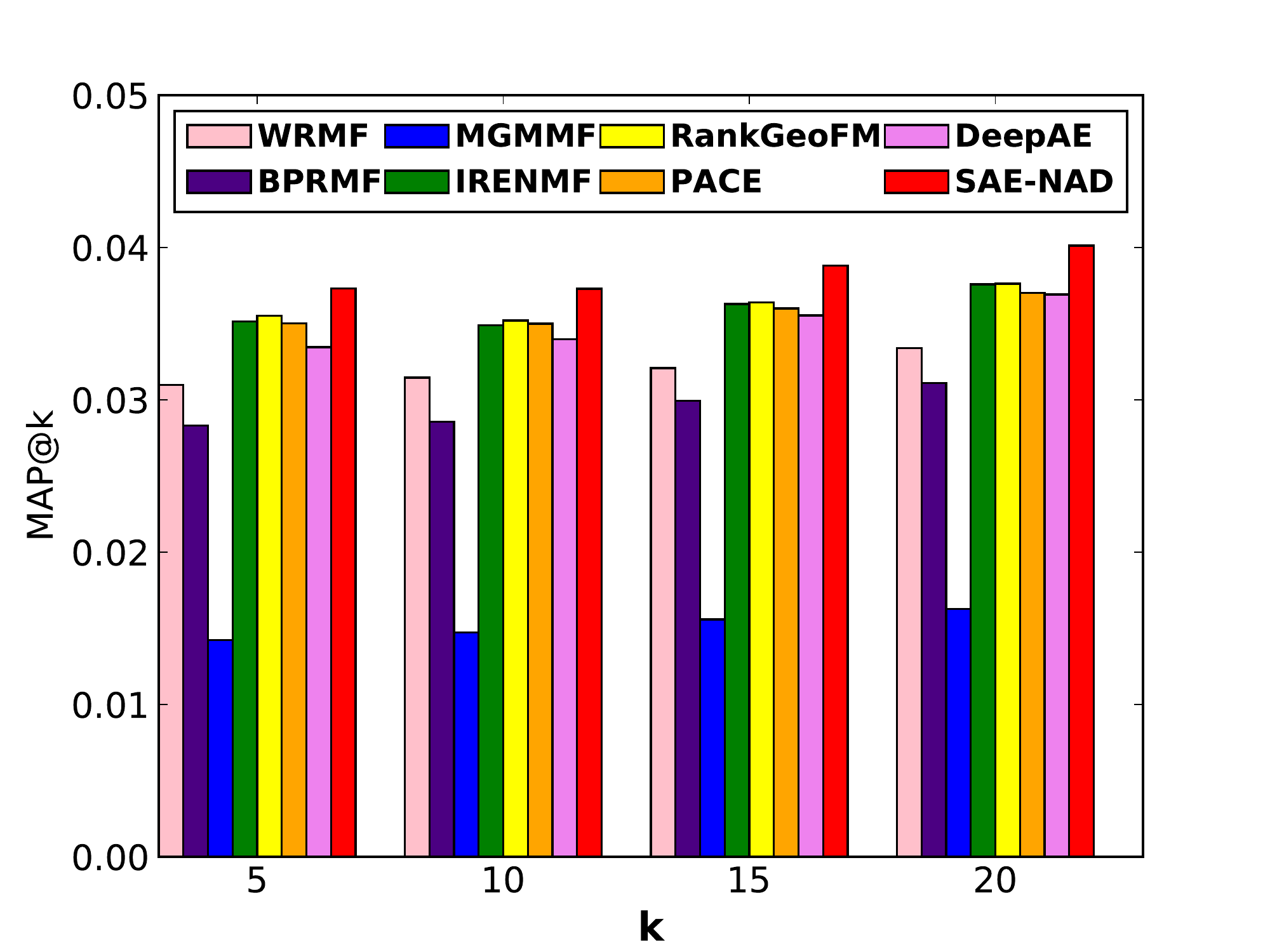}
        \caption{\label{fig:map_Yelp}MAP@k on Yelp}
    \end{subfigure}
    \caption{\label{fig:comparison_Yelp}The comparison of performance on Yelp.}
\vspace{-0.3cm}
\end{figure*}

\subsection{Parameter Settings}
In the experiments, the latent dimension of all the models is set to 50. The dimension of the importance vector $ d_{a} $ and the geographical correlation level $ \gamma $ are selected by grid search, which are set to 20 and 60, respectively. The parameters of the weighting scheme $ \alpha $ and $ \epsilon  $ are set to 2.0 and 1e-5, respectively. The gradient descent parameters, learning rate and regularization $ \lambda $, are set to 0.001 and 0.001, respectively.  $ a_{1} $-$ a_{3} $ are set as the $ tanh $ function, $ a_{4} $ is set to the $ sigmoid $ function. The batch size is set to 256. On the Gowalla dataset, we set the network architecture as $ [N, 500, 50, 500, N] $; otherwise, the network architecture is set as $ [N, 200, 50, 200, N] $. In addition, Dropout is used except for the first and last layer, where the Dropout probability is set to 0.5. Our model is implemented with PyTorch\footnote{https://pytorch.org/} running on GPU machines of Nvidia GeForce GTX 1080 Ti\footnote{Code is available at https://github.com/allenjack/SAE-NAD}.

For other baseline methods, following parameter settings achieve relatively good performance. \textit{DeepAE} adopts the same network architecture and weighted loss function with the proposed model. \textit{PACE} also uses the same network architecture (except for the hidden dimension) and parameters with the original paper. For \textit{RankGeoFM}, the number of the nearest neighbors is set to 300, the regularization radius $ C $ is set to 1.0, the regularization balance $ \alpha $ is set to 0.2, and the ranking margin $ \epsilon $ is set to 0.3 on all datasets. As for \textit{IRENMF}, $ \lambda_1 $, $ \lambda_2 $ and $ \lambda_3 $ are set to 0.015, 0.015 and 1, respectively; the instance weighting parameter $ \alpha $ is set to 0.6; as a preprocessing step, the model uses the k-means algorithm to cluster locations into 100 groups and the number of the nearest neighbors for each location is set to 10. For \textit{MGMMF}, the $ \alpha $ and $ \beta $ of the Poisson Factor Model are set to 20 and 0.2, respectively; $ \alpha $, $ \theta $ and the distance threshold $ d $ of the Multi-center Gaussian Model are set to 0.2, 0.02 and 15. \textit{WRMF} adopts the same weighting scheme as the proposed model.

\subsection{Performance Comparison}
The performance comparison of our model and baseline models are shown in Figure \ref{fig:comparison_Gowalla}, \ref{fig:comparison_Foursquare} and \ref{fig:comparison_Yelp}. 

\textbf{Observations about our model}. First, our proposed model--SAE-NAD achieves the best performance on three datasets with all evaluation metrics, which illustrates the superiority of our model. Second, SAE-NAD outperforms PACE, one possible reason is that PACE models the important geographical influence by a context graph, which does not explicitly model the user reachability to unvisited POIs. Instead, SAE-NAD directly captures the geographical influence between checked-in POIs and unvisited POIs through the neighbor-aware decoder. Third, SAE-NAD achieves better results than DeepAE, the major reason is that DeepAE only applies a multi-layer perceptron to model the check-in data without considering other context information in the check-in records. Fourth, SAE-NAD outperforms RankGeoFM and IRENMF. Although these two methods effectively incorporate geographical influence into a ranking model and an MF model, respectively, they still apply the inner product to predict users' preferences on POIs, which cannot sufficiently capture the non-linear interactions between users and POIs. On the other hand, SAE-NAD adopts a deep neural structure with non-linear activation functions to model the complex interactions in the user check-in data. Fifth, although MGMMF models the geographical influence effectively, it is not good at capturing user preferences from implicit feedback. Nevertheless, SAE-NAD encodes user's check-in frequencies into the weighting scheme, which indicates the confidence of users' preferences. Sixth, SAE-NAD outperforms BPRMF, because BPRMF only learns the pair-wise ranking of locations based on user preferences, it does not incorporate the context information such as spatial information of POIs. On the contrary, SAE-NAD integrates the geographical influence to further improve the performance. Besides, unlike existing methods that do not deeply explore the implicitness of users' preferences on checked-in POIs, SAE-NAD assigns an importance vector to each checked-in POI to characterize the user preference in multiple aspects.

\textbf{Other observations}. First, PACE outperforms all other baseline methods, since its neural embedding part models the user-POI interactions through the implicit feedback data. In the meanwhile, the context graph incorporates the context knowledge from the unlabeled data. Second, RankGeoFM and IRENMF both perform relatively well, which confirms the results reported in \cite{DBLP:journals/pvldb/LiuPCY17}. Third, although DeepAE applies a deep neural structure with weighted loss for implicit feedback, it still does not achieve better results than RankGeoFM and IRENMF. The reason is that DeepAE does not adopt the geographical information which is distinct for POI recommendation. But DeepAE performs better than WRMF and BPR, which may confirm that a deep network structure with non-linear activation functions can capture more sophisticated relations. Fourth, both WRMF and BPRMF are superior to MGMMF, one possible reason is that MGMMF is based on the probabilistic factor model, which models user check-in frequencies directly, instead of modeling user preferences on POIs. On the other hand, WRMF and BPRMF are designed for implicit feedback. WRMF not only considers the observed check-ins, but also gives a small confidence to all unvisited locations. On the other hand, BPRMF leverages location pairs as training data and optimize for correctly ranking location pairs.

\begin{table}[ht]
\centering
\caption{\label{tab:compare_components}The performance of the self-attentive encoder and neighbor-aware decoder on Gowalla, Foursquare, and Yelp.}
\begin{tabular}{ |c|c|c|c| }
 \hline
 \textit{Gowalla} & P@10 & R@10 & MAP@10 \\
 \hline
 WAE & 0.05599 & 0.13819 & 0.06728 \\ 
 \hline
 SAE-WAE & 0.06039 & 0.14808 & 0.07257 \\ 
 \hline
 NAD-WAE & 0.07029 & 0.17915 & 0.08699 \\ 
 \hline
\end{tabular}
\begin{tabular}{ |c|c|c|c| }
 \hline
 \textit{Foursquare} & P@10 & R@10 & MAP@10 \\
 \hline
 WAE & 0.05961 & 0.11134 & 0.05632 \\ 
 \hline
 SAE-WAE & 0.06346 & 0.11813 & 0.06054 \\ 
 \hline
 NAD-WAE & 0.06598 & 0.12546 & 0.06333 \\ 
 \hline
\end{tabular}
\begin{tabular}{ |c|c|c|c| }
 \hline
 \textit{Yelp} & P@10 & R@10 & MAP@10 \\
 \hline
 WAE & 0.03764 & 0.07386 & 0.03198 \\ 
 \hline
 SAE-WAE & 0.03951 & 0.07586 & 0.03307 \\ 
 \hline
 NAD-WAE & 0.04115 & 0.08016 & 0.03402 \\ 
 \hline
\end{tabular}
%\vspace{-0.3cm}
\end{table}

\subsection{Impacts of Self-Attentive Encoder and Neighbor-Aware Decoder}
The self-attentive encoder and neighbor-aware decoder are two important components of the proposed model. To verify the performance of each component, we solely evaluate each component in the weighted stacked autoencoder (section \ref{subsec:weighted_AE}). Here, we denote the stacked autoencoder with the weighted loss as \textit{WAE} (equals to DeepAE), the self-attentive encoder (SAE) with WAE as \textit{SAE-WAE}, and the neighbor-aware decoder (NAD) with WAE as \textit{NAD-WAE}. The performance is shown in Table \ref{tab:compare_components}.

The results in Table \ref{tab:compare_components} exhibit the effectiveness of the individual component of the proposed model. There are several observations: (1) The autoencoder with the weighted loss (WAE) achieves a reasonably good result, which even better than some baseline methods that incorporating the geographical influence. This illustrates that the frequency of the implicit feedback is a significant point to reveal user preferences. (2) By adopting the self-attention mechanism, SAE-WAE outperforms WAE on three datasets. The reason is that the self-attentive encoder attends the POIs that are more representative to reflect user preferences, leading to more personalized and effective user hidden representations. (3) NAD-WAE achieves better performance than SAE-WAE and WAE on three datasets. The reason why NAD-WAE performs better is that NAD-WAE captures the correlations between checked-in POIs and unvisited POIs, and applies these effects to the last layer of the decoder which directly determines the model output. The results further confirm that modeling geographical influence is essential for POI recommendation.

%\subsubsection{Sensitivity of Parameters}
\subsection{Sensitivity of Parameters}

\begin{figure}[t!]
    \centering
    \begin{subfigure}[t]{0.25\textwidth}
        \centering
        \includegraphics[width=\linewidth]{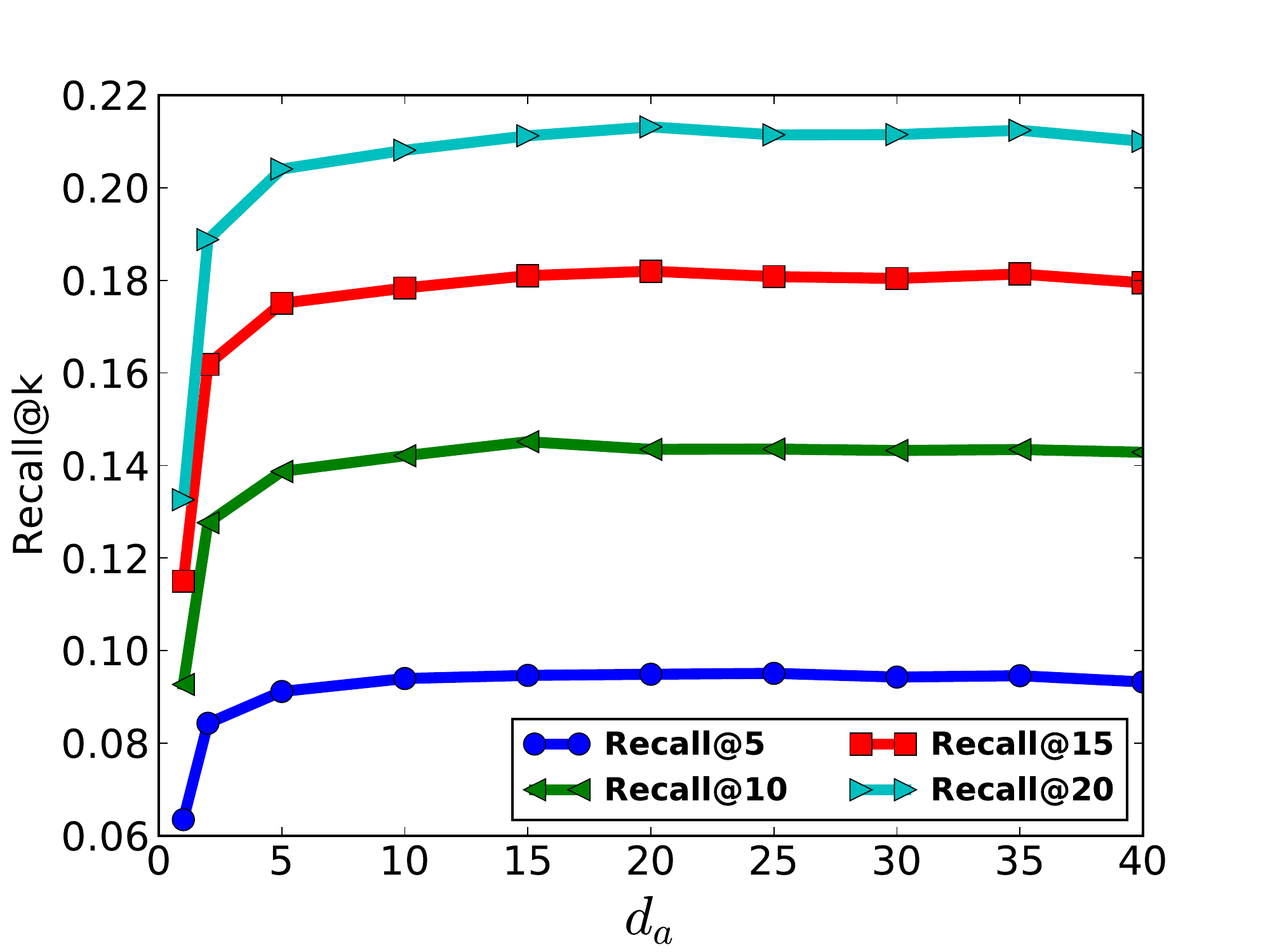}
        \caption{\label{fig:Gowalla_da_var}$ d_{a} $ on Gowalla.}
    \end{subfigure}%
    \begin{subfigure}[t]{0.25\textwidth}
        \centering
        \includegraphics[width=\linewidth]{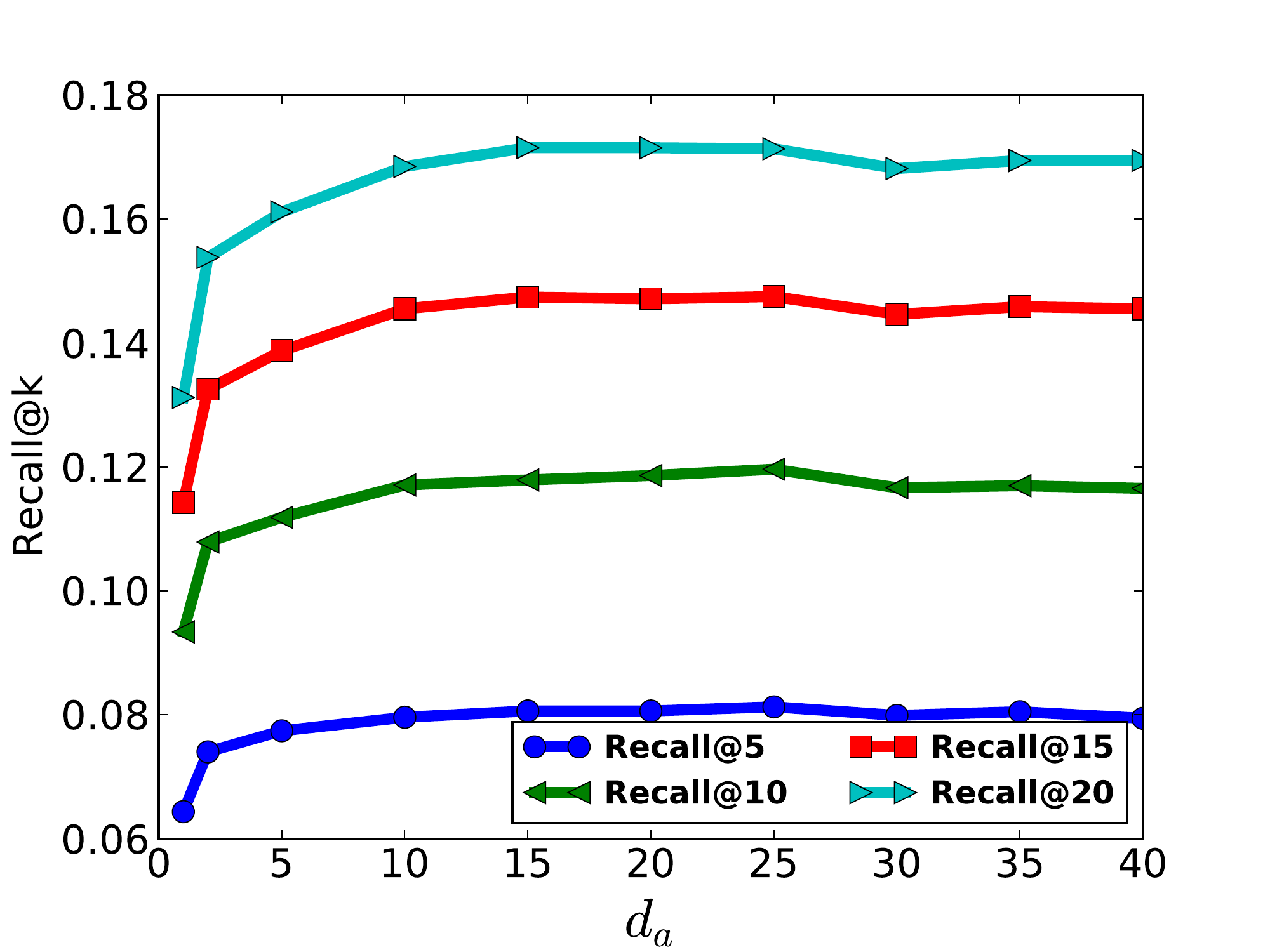}
        \caption{\label{fig:Foursquare_da_var}$ d_{a} $ on Foursquare.}
    \end{subfigure}
    \caption{\label{fig:da_var}The effect of $ d_{a} $.}
\vspace{-0.3cm}
\end{figure}

\begin{figure}[t!]
    \centering
    \begin{subfigure}[t]{0.25\textwidth}
        \centering
        \includegraphics[width=\linewidth]{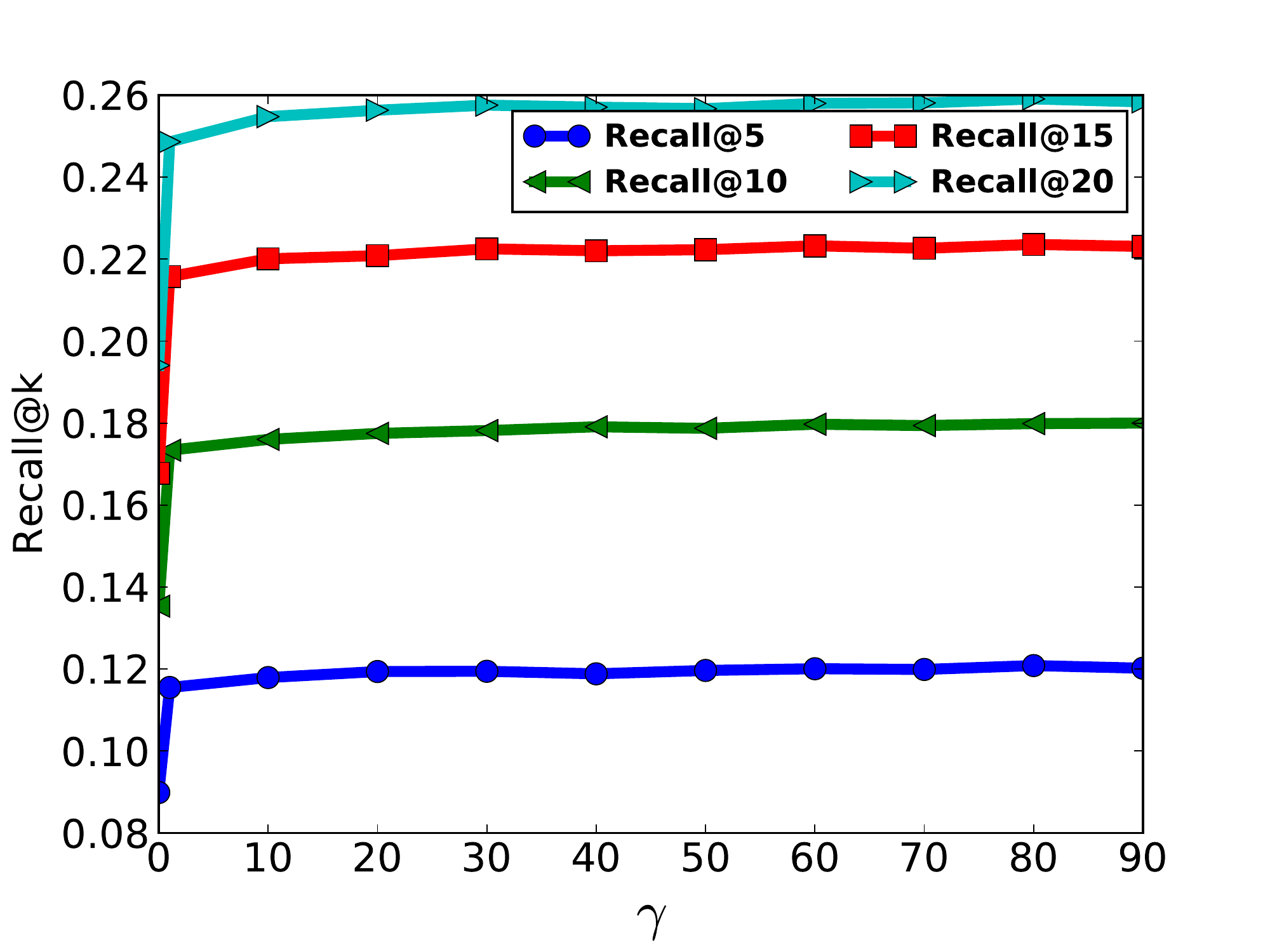}
        \caption{\label{fig:Gowalla_gamma_var}$ \gamma $ on Gowalla.}
    \end{subfigure}%
    \begin{subfigure}[t]{0.25\textwidth}
        \centering
        \includegraphics[width=\linewidth]{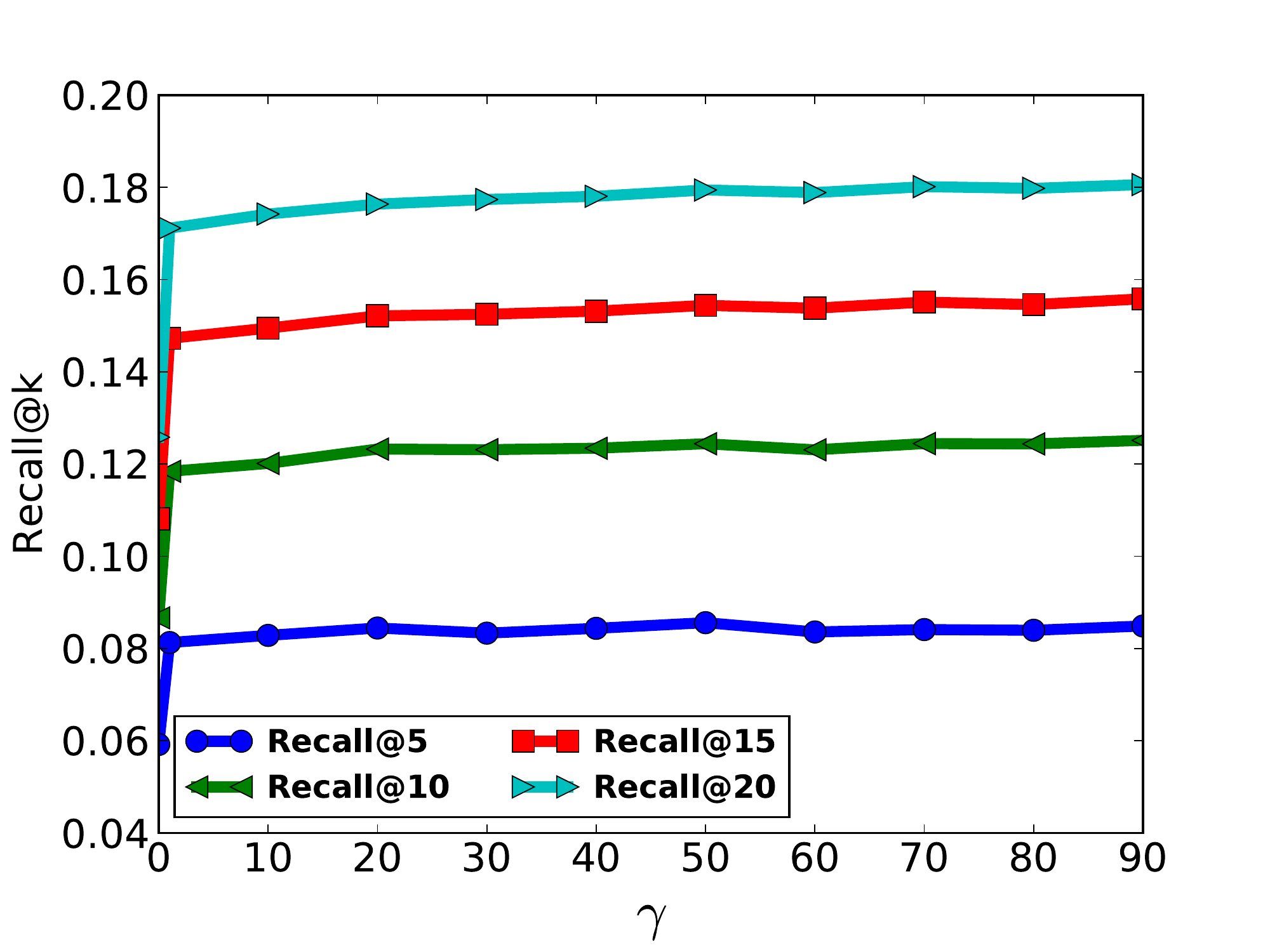}
        \caption{\label{fig:Foursquare_gamma_var}$ \gamma $ on Foursquare.}
    \end{subfigure}
    \caption{\label{fig:gamma_var}The effect of $ \gamma $.}
\vspace{-0.3cm}
\end{figure}

In the proposed model, two hyper-parameters are critical for performance improvement: the number of attention aspects $ d_{a} $ in the self-attentive encoder (section \ref{subsec:self-attention}) and the geographical correlation level $ \gamma $ of POIs in the neighbor-aware decoder (section \ref{subsec:neighbor_influence}). The effects of these two parameters are shown in Figure \ref{fig:da_var} and \ref{fig:gamma_var}. Due to the space limit, we only present the effects on Gowalla and Foursquare datasets, the parameter effects on Yelp dataset have similar trends.

The variation of $ d_{a} $ is shown in Figure \ref{fig:da_var}. We can observe that a single importance value from the attention layer is not sufficient to express the complex human sentiment on checked-in POIs. By assigning an importance vector to each checked-in POI, the user preference on those visited POIs can be captured from different aspects. With the increase of $ d_{a} $, the model performance largely improves and becomes steady.

The variation of $ \gamma $ is shown in Figure \ref{fig:gamma_var}. From the figure, we can observe that when $ \gamma = 0 $ the model does not consider the distance between POIs, leading to poor results. This also testifies the significance of geographical influence in POI recommendation. The larger value of $ \gamma $ strengthens the correlated level between two certain POIs, where neighbors of checked-in POIs will make a big difference in the inference of users' preferences.

\section{Conclusion}
In this paper, we proposed an autoencoder-based model for POI recommendation, which consists of a self-attentive encoder and a neighbor-aware decoder. In particular, the self-attentive encoder was used to adaptively discriminate the degree of user preference on each checked-in POI, by assigning an importance score vector. The neighbor-aware decoder was adopted to model the geographical influence checked-in POIs exerted on unvisited POIs, which differentiates the user reachability on unvisited POIs. Experimental results on three real-world datasets clearly validated the improvements of our model over many state-of-the-art baseline methods.

\bibliographystyle{ACM-Reference-Format}
\bibliography{weight_AE} 

\end{document}